\documentclass[10pt,conference]{IEEEtran}

\usepackage{cite}
\usepackage{amsmath,amssymb,amsfonts}
\usepackage{graphicx}
\usepackage{textcomp}
\usepackage{xcolor}
\usepackage[hyphens]{url}
\usepackage{fancyhdr}
\usepackage{hyperref}
\usepackage{cleveref}
\usepackage{booktabs}
\crefname{section}{§}{§}
\Crefname{section}{§}{§}
\usepackage{hhline}
\usepackage{booktabs}
\usepackage{multirow}
\usepackage[table]{xcolor}

\usepackage{algpseudocode}
\usepackage{enumitem}
\usepackage[ruled,linesnumbered,vlined]{algorithm2e}

\usepackage{multirow}
\usepackage{array}

\usepackage{subcaption}

\newcommand{\hpcayear}{2027}

\usepackage{xcolor}

\usepackage{comment}

\title{Energy-Efficient LLM Serving via Disaggregated Attention--FFN and Flexible Frequency Scaling}

\fancypagestyle{camerareadyfirstpage}{%
  \fancyhead{}
  
  \fancyhead[C]{
    \ifdefined\aeopen
    \parbox[][12mm][t]{13.5cm}{\hpcayear{} IEEE International Symposium on High-Performance Computer Architecture (HPCA)}    
    \else
      \ifdefined\aereviewed
      \parbox[][12mm][t]{13.5cm}{\hpcayear{} IEEE International Symposium on High-Performance Computer Architecture (HPCA)}
      \else
      \ifdefined\aereproduced
      \parbox[][12mm][t]{13.5cm}{\hpcayear{} IEEE International Symposium on High-Performance Computer Architecture (HPCA)}
      \else
      \parbox[][0mm][t]{13.5cm}{\hpcayear{} IEEE International Symposium on High-Performance Computer Architecture (HPCA)}
    \fi 
    \fi 
    \fi 
    \ifdefined\aeopen 
      \includegraphics[width=12mm,height=12mm]{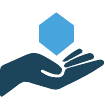}
    \fi 
    \ifdefined\aereviewed
      \includegraphics[width=12mm,height=12mm]{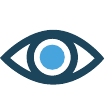}
    \fi 
    \ifdefined\aereproduced
      \includegraphics[width=12mm,height=12mm]{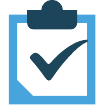}
    \fi
  }
  \fancyfoot[C]{}
}
\newcommand{\AFlex}{AFlex~}
\newcommand{\AFlexNaN}{AFlex}

\begin{document}

\author{
\IEEEauthorblockN{
Cunchen Hu\IEEEauthorrefmark{1},
Liangliang Xu\IEEEauthorrefmark{3},
Tian Liu\IEEEauthorrefmark{2},
Min Lyu\IEEEauthorrefmark{2},
Yongkun Li\IEEEauthorrefmark{2},
Sa Wang\IEEEauthorrefmark{4},\\
Shuo Quan\IEEEauthorrefmark{1},
Yanan Yang\IEEEauthorrefmark{1},
Wenda Tang\IEEEauthorrefmark{1},
Yiduo Wang\IEEEauthorrefmark{1},
Fu Yu\IEEEauthorrefmark{1},
Jie Wu\IEEEauthorrefmark{1}
}
\IEEEauthorblockA{
\IEEEauthorrefmark{1}
China Telecom Cloud Computing Research Institute\\
\IEEEauthorrefmark{3}
Xidian University\\
\IEEEauthorrefmark{2}
University of Science and Technology of China\\
\IEEEauthorrefmark{4}
SKLP, ICT, CAS
}
}

\maketitle

\begingroup
\renewcommand{\thefootnote}{\(\dagger\)}
\footnotetext{Liangliang Xu is the corresponding author.}
\endgroup

\ifdefined\hpcacameraready 
  \thispagestyle{camerareadyfirstpage}
  \pagestyle{empty}
\else
  \thispagestyle{plain}
  \pagestyle{plain}
\fi

\newcommand{\hpcaheight}{0mm}
\ifdefined\eaopen
\renewcommand{\hpcaheight}{12mm}
\fi


\begin{abstract}
Large language model (LLM) serving spans diverse applications with stringent service-level objectives (SLOs), often requiring GPUs to run at maximum frequencies and increasing energy consumption.
Existing energy-management approaches adapt GPU frequencies only at the request or inference-phase level, overlooking operator-level differences in frequency sensitivity between Attention and feed-forward networks (FFNs).
We find that the energy-optimal frequencies of Attention and FFN (A/F) differ and vary with the inference phase, workload, and system configurations. 
However, runtime variability and independent A/F frequency control
create a large search space and high communication overhead.
To address these challenges, we present AFlex, a framework that jointly optimizes resource provisioning and GPU frequency scaling for disaggregated A/F serving.
AFlex introduces a global scheduler and a local operator-level dynamic voltage and frequency scaling (DVFS) controller to determine A/F resource allocations and frequencies. It further introduces an interleaved A/F pipeline with dynamic microbatch depth and adaptive request batching to reduce pipeline bubbles.
We implement AFlex in SGLang and evaluate it on NVIDIA A800 GPUs using Qwen3-32B and Mixtral-8$\times$7B under production Conversation and Coding traces. 
\AFlex reduces energy per token by up to 49\% over state-of-the-art disaggregated serving and 48\% over frequency-scaling systems while satisfying TTFT and TPOT SLOs.
\end{abstract}

\section{Introduction}
\label{sec:introduction}


Large language model (LLM) serving has become a major GPU workload in modern datacenters~\cite{copilot, peng2023study, openai2023chatgpt, gemini-context-caching, quinn2025accelerating}. At scale, it consumes substantial GPU resources and energy while satisfying stringent service-level objectives (SLOs), including time to first token (TTFT) and time per output token (TPOT). Modern GPUs support dynamic voltage and frequency scaling (DVFS), enabling systems to trade performance for power savings. 
Therefore, SLO-aware GPU frequency control is critical for reducing serving energy.

LLM inference consists of prefill and decode (P/D) phases with distinct execution characteristics, and recent work~\cite{tetriserve-2024,hu2024memserve,zhong2024distserve} disaggregates the two phases to enable specialized resource provisioning and reduce cross-phase interference.
To reduce energy consumption under strict latency SLOs, prior work has explored DVFS control for LLM serving.
DynamoLLM~\cite{stojkovic2025dynamollm} jointly optimizes instance count, model parallelism, and GPU frequency under SLO constraints, while throttLL’eM~\cite{kakolyris2025throttll} predicts GPU frequencies from iteration-level dynamics.
GreenLLM~\cite{greenllm} and BiScale~\cite{basit2026biscale} further assign separate frequencies to prefill and decode phases based on their distinct execution characteristics. These systems demonstrate that DVFS can reduce serving energy while satisfying SLO constraints. 
However, phase-level granularity control still treats each phase as a single frequency domain, overlooking operator-level heterogeneity, particularly between Attention and feed-forward networks (FFNs).

Attention and FFN, the two dominant operators in LLM inference, exhibit distinct compute and memory characteristics, while their relative execution times vary across serving phases, workloads, and configurations. 
During prefill, both operators are dominated by matrix multiplications but scale differently with sequence length. Batch size and tensor parallelism (TP) degree further change matrix shapes and arithmetic intensity, which can shift each operator between compute-bound and memory-bound regimes. During decode, 
Attention is often memory-bound due to intensive accesses to the growing key-value (KV) cache,
whereas FFN is memory-bound at low batch sizes but becomes more compute-bound and frequency-sensitive with larger batches. 
These differences lead to distinct DVFS responses: higher GPU frequencies substantially accelerate compute-bound execution but provide limited benefits when memory access dominates. 
As a result, no single frequency consistently optimizes the latency--energy tradeoff across operators, even within the same P/D phase.


However, exploiting operator-level heterogeneity through Attention--FFN (A/F) disaggregation (AFD) and A/F-aware DVFS introduces two challenges.
(1) \textit{Finding energy-efficient A/F configurations at modest search cost is non-trivial.} Under a shared resource budget, satisfying the TTFT and TPOT SLOs requires coordinated P/D provisioning and, within each phase, joint A/F optimization for pipeline balance. Independent optimization at either level can waste resources or violate SLOs. Yet selecting A/F provisioning, TP layouts, and GPU frequencies creates a large combinatorial search space whose optimum shifts with the workload. Joint A/F scheduling must therefore adapt these configurations while coordinating cross-phase resources and allocating intra-phase latency slack.
(2) \textit{Communication and transition overheads in AFD may outweigh the benefits of fine-grained control.}
AFD requires hidden-state transfers at each transformer layer, incurring non-negligible communication overhead on the inference-critical path. Moreover, dynamically adapting A/F resource configurations and GPU frequencies introduces transition costs at different timescales. Without careful control, these communication and adaptation overheads can offset the energy savings from dynamic provisioning and frequency scaling. 
The system must therefore minimize per-layer communication and DVFS switching overhead while decoupling slow resource reconfiguration from the inference critical path.

To address these challenges, we present \AFlexNaN, the first LLM serving framework that combines disaggregated \textbf{\underline{A}}ttention and \textbf{\underline{F}}FN execution with f\textbf{\underline{lex}}ible per-operator frequency scaling.
\AFlex combines a two-level control plane with a pipelined A/F data plane.
(1) The control plane comprises a Global Scheduler and a Local DVFS Controller. Using offline profiles, the Global Scheduler periodically solves an Integer Linear Program (ILP) to select A/F pairs, TP degrees, and base frequencies under resource, throughput, and latency constraints. The Local DVFS Controller adapts these frequencies at runtime, using one Prefill Attention  / Prefill FFN (PA/PF) pair per prefill batch and adaptive Decode Attention / Decode FFN (DA/DF) windows to amortize switching costs.
(2) The data plane employs an interleaved A/F pipeline that launches next-layer Attention once its FFN completes, overlapping hidden-state transfers with computation. It dynamically selects the microbatch depth and request batch size to balance A/F execution and reduce pipeline bubbles.

We implement \AFlex atop SGLang~\cite{sglang} and evaluate it with Qwen3-32B~\cite{qwen3technicalreport} and Mixtral-8$\times$7B~\cite{Mixtral} on two servers, each with eight NVIDIA A800 GPUs, and scale to four servers for the scalability study. Our workloads include two production traces for conversation and coding and four controlled workloads spanning input and output lengths. We compare \AFlex against P/D-colocated SGLang, DVFS-enabled P/D-colocated DynamoLLM~\cite{stojkovic2025dynamollm}, P/D-disaggregated DistServe~\cite{zhong2024distserve}, and DVFS-enabled P/D-disaggregated BiScale~\cite{basit2026biscale}. We analyze component contributions, scalability, and system overhead. \AFlex reduces energy per token by up to 49\% over state-of-the-art disaggregated serving and 48\% over frequency-scaling systems while satisfying TTFT and TPOT SLOs.

In summary, this paper makes the following contributions:
\begin{itemize}[leftmargin=*,itemsep=2pt,parsep=0pt, topsep=2pt]
    \item We identify and systematically characterize operator-level frequency heterogeneity in LLM serving, revealing opportunities for A/F-aware energy optimization across inference phases, workloads, and system configurations.

    \item We develop a two-level control plane that combines global A/F resource provisioning with local per-operator DVFS adaptation to minimize serving energy under resource, throughput, and latency constraints.

    \item We design an interleaved A/F data plane that overlaps hidden-state transfers with computation and dynamically selects the microbatch depth and adaptive request batching to balance concurrent stages and reduce pipeline bubbles.
    
    \item We evaluate \AFlex on a multi-node GPU cluster using production-level traces, showing substantial energy reductions under SLO constraints compared with state-of-the-art disaggregated serving and frequency-scaling systems.
\end{itemize}
We will release the source code in the final version.

\section{Background}
\label{sec:background}

\noindent\textbf{LLM Serving.}
Generative LLM serving consists of a compute-intensive \emph{prefill} phase that processes prompt tokens in parallel and populates the KV cache, followed by a sequential, memory-bound \emph{decode} phase that generates tokens using cached states~\cite{yu2022orca,tensorrt-llm}. Serving performance is measured by TTFT and TPOT, with interactive applications imposing SLOs on one or both metrics. These metrics vary with prompt and output lengths, request concurrency, batch size, TP, and KV cache utilization, complicating simultaneous SLO satisfaction. Within each phase, Attention and FFN also exhibit distinct resource demands: Attention is sensitive to sequence length, KV cache access, and data movement, whereas FFN relies primarily on compute-intensive matrix multiplications. Consequently, the two operators respond differently to resource allocation and system configurations.

\noindent\textbf{Phase and Operator Disaggregation.}
Many LLM serving systems colocate prefill and decode on the same GPUs and interleave their execution through continuous batching~\cite{vllm-sosp23,agrawal2023sarathi,fastserve-arxiv23,yu2022orca}. Although this improves GPU utilization, long prefill requests can delay ongoing decode iterations, increasing per-token latency and complicating simultaneous TTFT and TPOT SLO satisfaction. Chunked prefill mitigates this interference but requires workload-dependent tuning~\cite{agrawal2024taming,agrawal2023sarathi}. 
P/D disaggregation instead assigns the two phases to separate GPU pools, enabling phase-specific batching, provisioning, and scaling to better balance TTFT and TPOT~\cite{zhong2024distserve,patel2023splitwise,tetriserve-2024}.
However, P/D disaggregation remains coarse-grained because each phase still executes alternating Attention and FFN operators with distinct compute, memory, and resource demands. AFD further exposes this heterogeneity by assigning the two operators to separate GPU groups and transferring hidden states across layer boundaries~\cite{zhu2025megascale,xiao2025xdeepserve}. This finer granularity enables operator-specific resource provisioning and improves performance scalability over P/D disaggregation.

\noindent\textbf{Energy Efficiency with DVFS.}
DVFS improves energy efficiency by trading GPU performance for lower power consumption. Prior work adjusts GPU frequency based on workload conditions or assigns separate frequencies to prefill and decode, but still applies the same frequency to Attention and FFN despite their different sensitivities~\cite{stojkovic2025dynamollm,basit2026biscale}.
FFN is typically more compute-sensitive, whereas Attention is often more memory-bound and benefits less from higher GPU frequencies. A frequency chosen for FFN may therefore waste energy on Attention, while one chosen for Attention may increase FFN latency. Moreover, the optimal frequency for each operator varies with workload and parallel configuration, and frequent frequency changes introduce non-negligible overhead. Therefore, exposing Attention and FFN as disaggregated execution and frequency-control domains, while coordinating their decisions under TTFT and TPOT SLOs, is important for energy-efficient LLM serving.
\section{Observations}
\label{sec:observation}

To understand the energy-efficiency properties of LLM serving with AFD, we characterize energy and performance on an eight-GPU NVIDIA A800-80GB server across GPU frequencies, TP degrees, batch sizes, and sequence lengths. We first analyze the ideal energy-savings opportunity under both homogeneous and heterogeneous A/F configurations. 
We further characterize A/F operators in terms of frequency sensitivity, energy efficiency under homogeneous provisioning, and configuration overhead.
Finally, we analyze the challenges of the scheduling algorithm and system design.

\begin{figure}[t!]
  \centering
  \centerline{\includegraphics[width=0.48\textwidth]{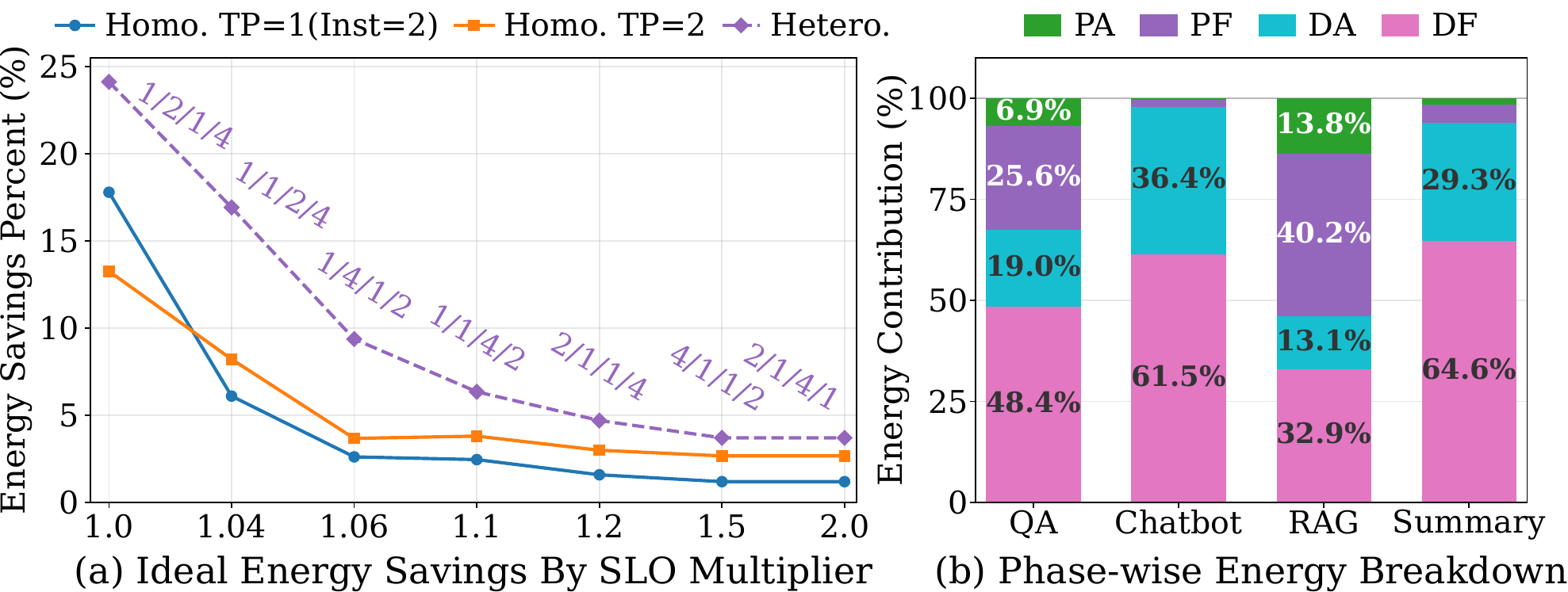}}
  \caption{Estimated ideal energy savings achieved by homogeneous and heterogeneous A/F provisioning across workloads.
 (a) Energy savings under homogeneous and heterogeneous provisioning. (b) Energy breakdown for QA (128/64), Chatbot (128/1024), RAG (4096/64), and Summary (4096/1024), where numbers denote input/output token lengths. Each tuple denotes the TP degrees of PA, PF, DA, and DF, respectively.}
  \label{fig:obs1}
\end{figure}



\subsection{Energy-Saving Opportunity of AFD}
\label{sec:obs-saving-potential}
To estimate the ideal energy-saving opportunity, we scan GPU frequencies across multiple latency SLOs, A/F provisioning configurations, and input/output workloads, using $frequency \times latency$ as a lightweight energy proxy. We set each SLO to $1.0\times$--$2.0\times$ the maximum-frequency latency 
and select the configuration with the lowest energy proxy value among those satisfying the SLO.

Fig.~\ref{fig:obs1}(a) shows substantial estimated
SLO-constrained energy-saving opportunities under both
homogeneous and heterogeneous provisioning.
Heterogeneous A/F provisioning creates the largest optimization
opportunity, reducing energy proxy by up to 48.1\% over homogeneous provisioning
under the same GPU budget.
Fig.~\ref{fig:obs1}(b) further shows that PA, PF, DA, and DF contribute markedly different fractions of the total energy proxy across input/output workloads.
Specifically, PA contributes 0.4--13.8\%, PF 1.7--40.2\%, DA 13.1--36.4\%, and
DF 32.9--64.6\% of total energy proxy. This variation suggests that A/F optimization
must jointly account for prefill and decode behavior across diverse
input/output workloads.

\noindent\textbf{Insight \#1.}
\textbf{AFD creates significant energy-saving opportunities:
heterogeneous provisioning yields greater savings than homogeneous provisioning,
with PA, PF, DA, and DF contributing differently across workloads.}

\subsection{A/F Frequency Sensitivity in Performance and Energy}
\label{sec:obs-misaligned-optima}
\begin{figure}[t!]
  \centering
  \centerline{\includegraphics[width=0.48\textwidth]{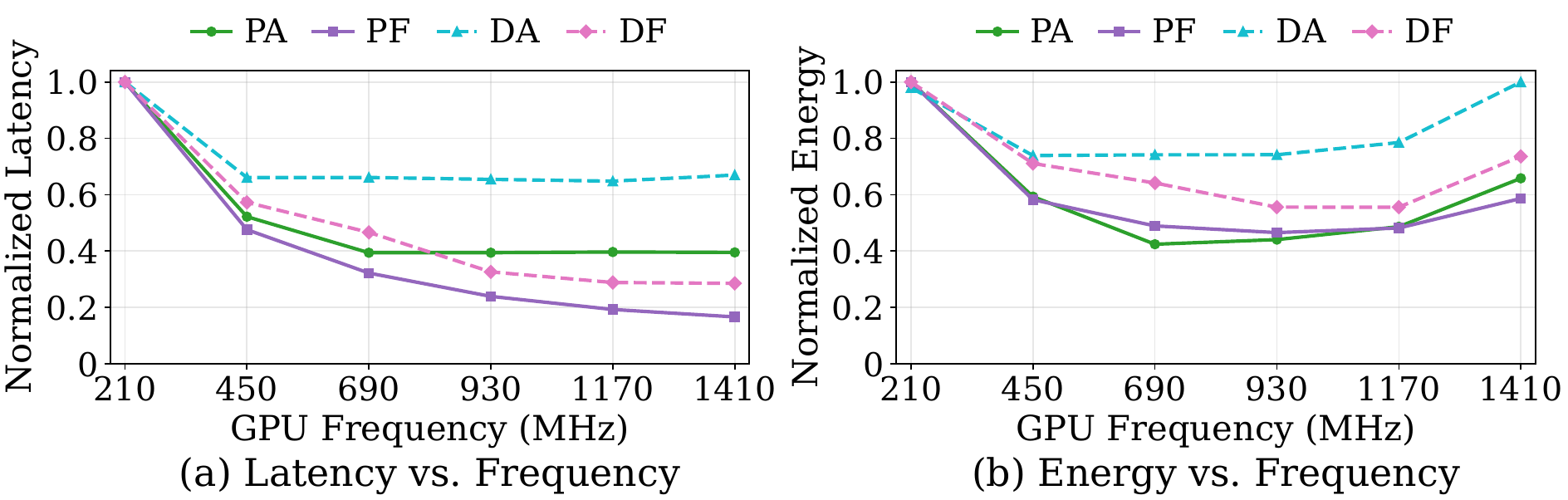}}
  \caption{A/F frequency sensitivity: (a) Normalized latency and (b) Energy across GPU frequencies.}
  \label{fig:obs2}
\end{figure}

We characterize the frequency sensitivity of Attention and FFN in terms of both latency and energy. By default, we set TP to 4 and the batch size to 64.
Fig.~\ref{fig:obs2}(a) shows that the two operators respond differently to the same GPU frequency change across the prefill and decode phases. In decode, increasing the GPU frequency from 210\,MHz to 1410\,MHz reduces Attention latency by 33.0\%, but reduces FFN latency by 71.5\%. In prefill, the same frequency increase reduces Attention and FFN latency by 60.5\% and 83.4\%, respectively.
Fig.~\ref{fig:obs2}(b) reveals a distinct U-shaped energy response to frequency scaling. In both prefill and decode, increasing frequency initially reduces energy by shortening execution time, but eventually increases energy as power overhead dominates. Moreover, Attention and FFN often reach their energy minima at different frequencies. Across the profiled configurations, Attention is energy-optimal at a lower frequency than FFN in 63.8\% of cases, with an average gap of 429\,MHz.
These results show that a single GPU frequency cannot consistently match both the latency sensitivity and energy optima of Attention and FFN.

\begin{figure}[t!]
  \centering
  \centerline{\includegraphics[width=0.48\textwidth]{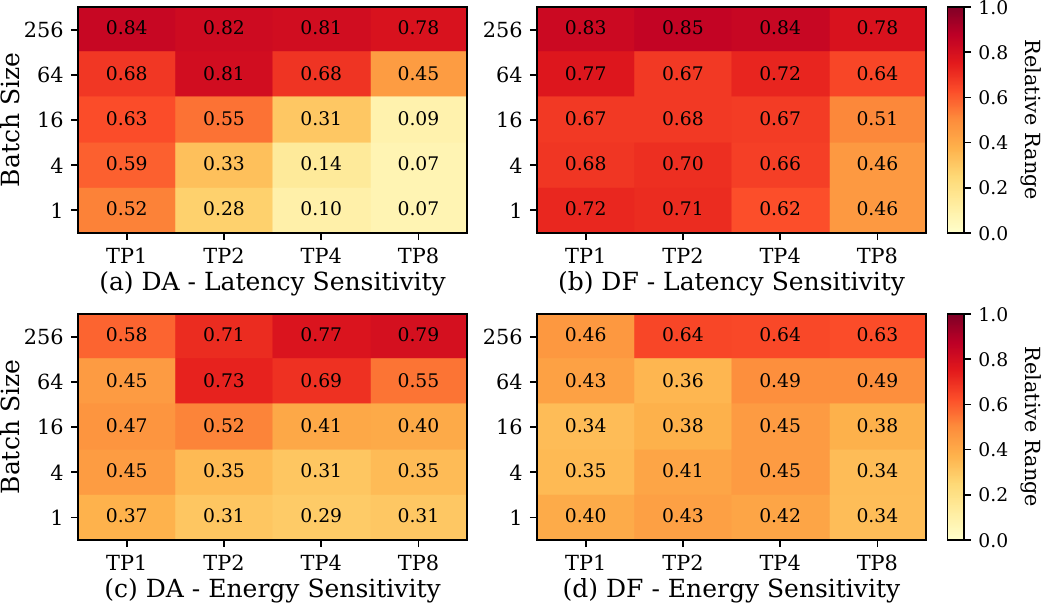}}
  \caption{
  Frequency sensitivity across decode batch sizes and TP degrees. Each cell reports operator sensitivity to GPU frequency as one minus the minimum-to-maximum latency or energy ratio; darker colors indicate higher sensitivity.
}

  \label{fig:obs3}
\end{figure}

\noindent\textbf{Insight \#2.}
\textbf{A/F operators prefer different GPU frequencies: their latency
sensitivities and energy-optimal points differ along U-shaped
energy--frequency curves, making a single shared frequency suboptimal.}

We further characterize the frequency sensitivity of Attention and FFN across
batch sizes and TP degrees. For each setting, we report the operator's
sensitivity to GPU frequency, measured as one minus the minimum-to-maximum
ratio of latency or energy across frequencies.
Fig.~\ref{fig:obs3} shows that the latency and energy sensitivity of both
operators vary substantially across batch sizes and TP degrees. Across settings,
the energy sensitivity ranges from 0.29 to 0.79 for Attention and from 0.34 to
0.64 for FFN. The corresponding latency sensitivity ranges from 0.07 to 0.84
for Attention and from 0.46 to 0.85 for FFN. These large and operator-dependent variations indicate that the optimal
energy--latency frequency is runtime-dependent, making a fixed per-operator
frequency suboptimal across batch sizes and TP degrees.

\noindent\textbf{Insight \#3.}
\textbf{A/F operators require dynamic GPU frequencies: their latency and energy optima shift across runtime configurations, including batch size and TP degree, making fixed per-operator frequencies suboptimal.
}

\begin{figure}[t!]
  \centering
  \centerline{\includegraphics[width=0.48\textwidth]{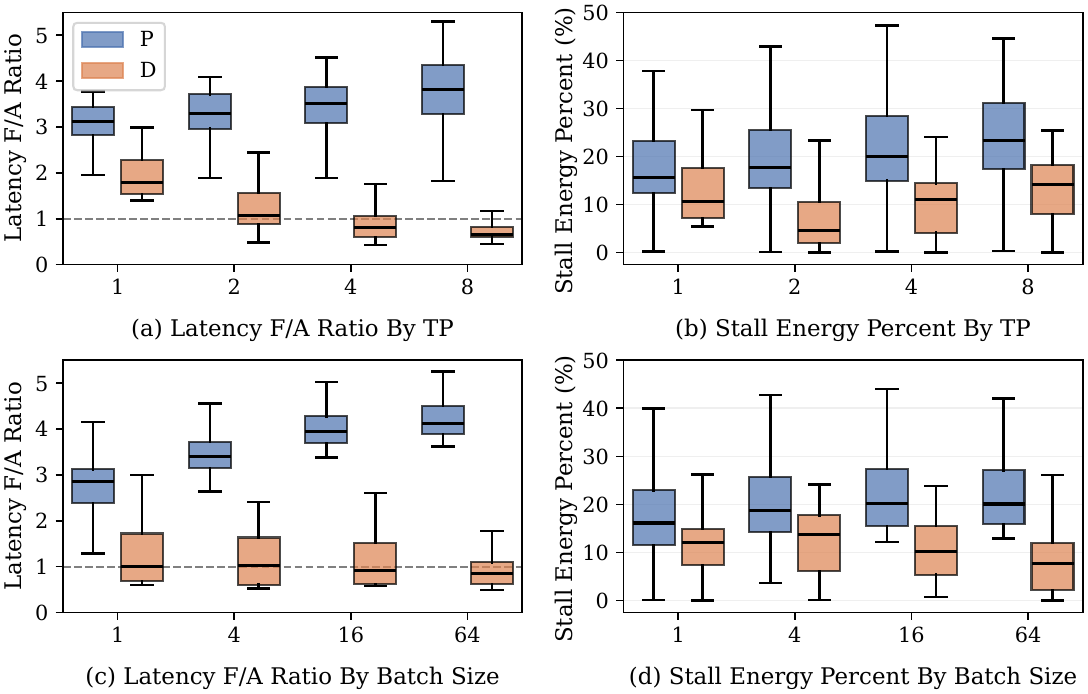}}
    \caption{A/F latency imbalance and energy loss under homogeneous provisioning
    across TP degrees and batch sizes.}
  \label{fig:obs4}
\end{figure}

\subsection{Energy Inefficiency under Homogeneous A/F Provisioning}
\label{sec:obs-dynamic-provisioning}
We characterize the A/F latency ratio and wasted energy under homogeneous A/F
provisioning across TP degrees and batch sizes.
Fig.~\ref{fig:obs4}(a) and (c) show that homogeneous provisioning cannot
maintain a stable A/F balance across TP degrees and batch sizes. The
FFN-to-Attention latency ratio ranges from 0.43 to 5.30, deviating substantially
from the balanced value of 1.
As a result, a fixed resource allocation can leave one side idle while the other
dominates step latency.
Fig.~\ref{fig:obs4}(b) and (d) show that this imbalance directly increases
energy consumption. The resulting bubble time accounts for 19.7--75.7\% of the
pipeline step, leading to 1.1--30.6\% energy loss under homogeneous
provisioning. Thus, adaptive provisioning is not only a latency optimization
but also necessary for SLO-constrained energy minimization.

\noindent\textbf{Insight \#4.}
\textbf{Homogeneous A/F provisioning is suboptimal: A/F balance shifts across TP degrees and batch sizes, causing pipeline bubbles and energy loss that require dynamic heterogeneous provisioning.}

\subsection{A/F Configuration Overhead}
\label{sec:obs-reconfig-overhead}
\begin{figure}[t!]
  \centering
  \centerline{\includegraphics[width=0.48\textwidth]{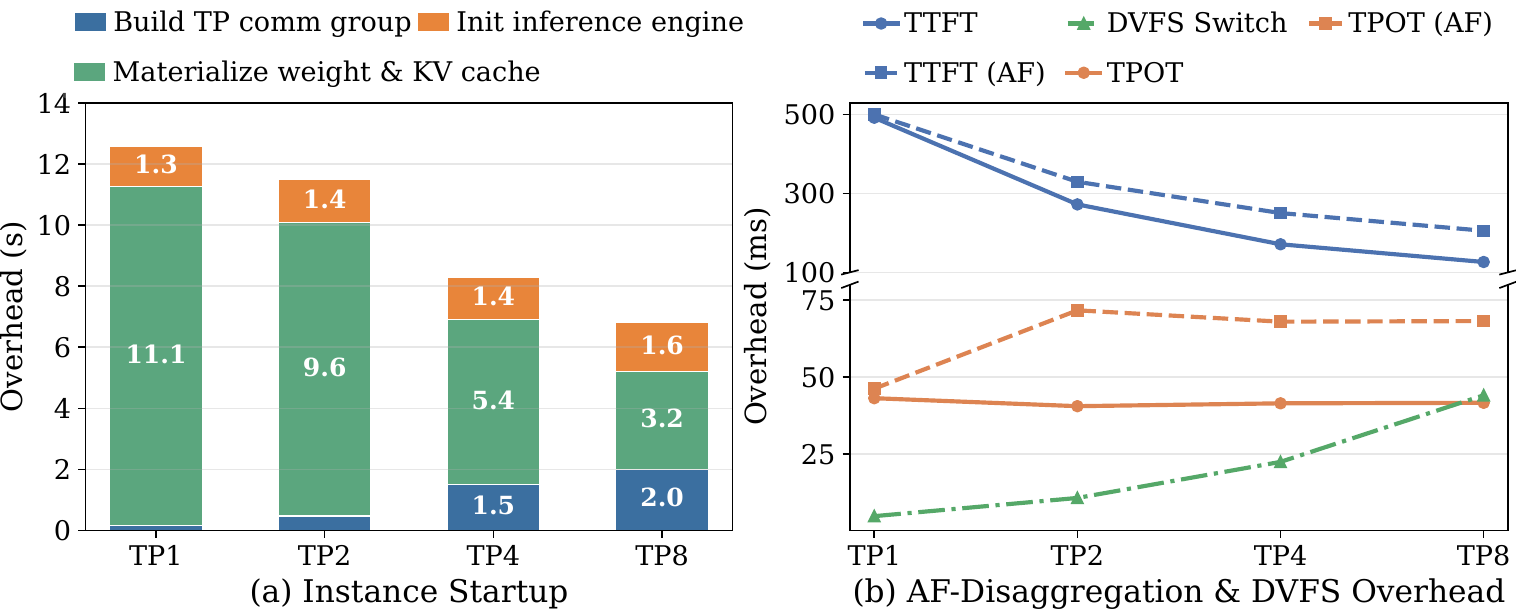}}

  \caption{A/F configuration overhead. (a) Instance startup overhead. (b) A/F transfer  \& DVFS switching overhead. }
  \label{fig:obs5}
\end{figure}

We characterize A/F configuration overhead from instance startup, hidden-state transfer, and DVFS switching.
Fig.~\ref{fig:obs5}(a) shows that A/F TP-layout reconfiguration introduces non-negligible overhead, including rebuilding TP communication groups, materializing weights and KV cache, and initializing the inference engine. Across TP degrees, the initialization time ranges from 12.6\,s at TP=1 to 6.8\,s at TP=8, decreasing as more GPUs participate in parallel initialization. 
Fig.~\ref{fig:obs5}(b) shows that GPU frequency switching and A/F hidden-state transfer both introduce millisecond-scale overhead. Updating GPU frequencies takes 4.7--44.1\,ms as the number of GPUs on the same host increases from 1 to 8, due to serialized clock updates in the NVIDIA driver. A/F hidden-state transfer adds 8--79\,ms in prefill and 3--32\,ms in decode. These costs are non-negligible relative to TTFT (126--492\,ms) and TPOT (41\,ms), indicating that switching and communication overhead must be amortized differently across phases.

\noindent\textbf{Insight \#5.}
\textbf{A/F configuration must be overhead-aware: seconds-scale TP-layout
reconfiguration discourages fine-grained layout changes, while millisecond-scale
DVFS switching and A/F communication must be amortized differently across
prefill and decode.}

\section{AFlex Design}
\label{sec:design}                                                                                           
\label{overview}

Guided by the insights from \cref{sec:observation}, we design \AFlexNaN, an operator-level disaggregated energy-management framework for LLM serving. 
\AFlex follows four design principles. (1) It is operator-aware, exposing PA, PF, DA, and DF as separate execution and frequency-control domains for finer-grained control than unified GPU frequency scaling.
(2) It is jointly optimized and SLO-aware, using one ILP over the four operator-phase pools to select energy-efficient configurations under TTFT and TPOT constraints.
(3) It separates resource planning from runtime control, using coarse-grained orchestration for SLO feasibility and runtime DVFS for per-operator slack recovery.
(4) It supports coordinated operator-phase reconfiguration under fluctuating loads, distinguishing lightweight frequency transitions from heavier TP adjustments to bound overheads.

To our knowledge, \AFlex is the first LLM serving framework to combine operator-disaggregated execution with independent GPU frequency scaling for Attention and FFN. Unlike existing energy-management frameworks~\cite{stojkovic2025dynamollm, kakolyris2025throttll, basit2026biscale}, \AFlex exploits operator-level DVFS sensitivity that varies across phases, sequence lengths, and batch sizes. It lowers the frequency of less sensitive operators, reducing energy without violating TTFT or TPOT constraints.

\begin{figure}[!t]
    \centering
  \centerline{\includegraphics[width=0.48\textwidth]{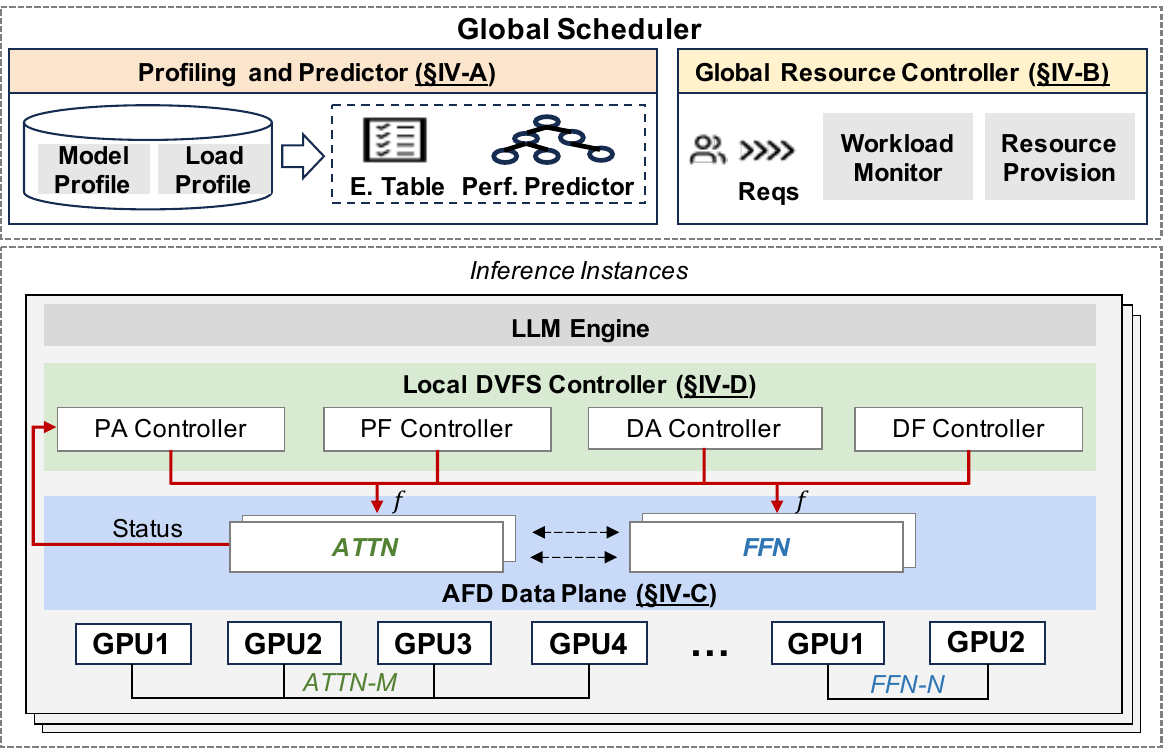}}
    \caption{The architecture of \AFlexNaN.}
    \label{fig:overview}
\end{figure}



\noindent\textbf{Architecture.}
Fig.~\ref{fig:overview} shows the \AFlex architecture, which comprises a global scheduler, an AFD data plane, and a local DVFS controller.
The global scheduler has two components.
First, the profiling and prediction module performs offline profiling and generates an energy table and a performance predictor.
Second, the global resource controller monitors workload changes and adjusts resource provisioning based on these models.
It periodically solves a joint ILP to determine the GPU allocation, TP degree, and baseline frequencies for each operator pool under SLO constraints.
The AFD data plane separates Attention and FFN execution into distinct operator pools.
The PA and PF pools serve the prefill phase under the TTFT constraint, whereas the DA and DF pools serve the decode phase under the TPOT constraint.
Within each phase, Attention and FFN units are connected through layerwise hidden-state transfers.
The local DVFS controller manages the PA, PF, DA, and DF controllers, which adjust per-operator frequencies based on runtime status and SLO constraints.


\subsection{Operator-Level Profiling and Prediction}
\label{sec:profiling}


To capture operator-specific characteristics, \AFlex profiles A/F operators in different configurations and trains lightweight predictors to estimate their latency and energy consumption.

\noindent\textbf{Profiling.} \AFlex collects these profiles offline on the target GPU platform across the P/D phase by sweeping TP degree, GPU frequency, batch size, and sequence length.
Each profile entry records the measured latency and GPU energy consumption of an Attention or FFN operator under a specific configuration.
Under tensor parallelism, operator latency is determined by the slowest rank in the TP group, while energy consumption is aggregated across all participating GPUs.
Because Attention and FFN execute in disaggregated resource pools, \AFlex separately profiles hidden-state transfers between the two pools to capture their communication overhead.

\noindent\textbf{Prediction.} \AFlex trains lightweight regression models on the offline profiles to predict operator latency and energy. For each operator, \AFlex builds separate latency and energy predictors that take GPU frequency, batch size, sequence length, and TP degree as input features.
We use gradient-boosted decision trees to capture non-linear feature interactions while keeping inference overhead low. These models guide the global resource controller and local DVFS controller in selecting energy-efficient configurations under SLO constraints.




\subsection{Global Resource Controller}
\label{sec:tier1}


The Global Resource Controller periodically adapts cluster configurations to workload drift through three components: a \textit{workload monitor} that tracks health and provisioning pressure, a \textit{resource orchestrator} that solves an ILP for the minimum-energy feasible configuration, and a \textit{runtime reconfiguration engine} that applies the configuration with minimal disruption.

\noindent\textbf{Workload Monitor.}
\label{sec:tier1:monitor}
The Workload Monitor tracks cluster health every 300\,s and triggers reconfiguration when any of the following four cases occur:
\begin{itemize}[leftmargin=*,itemsep=2pt,parsep=0pt, topsep=2pt]
    \item \underline{\textit{SLO violation:}} a non-negligible fraction of requests exceeds the TTFT or TPOT target, indicating that the current provisioning lacks sufficient effective capacity, especially when assigned GPUs already run at maximum frequency.
    
    \item \underline{\textit{A/F imbalance:}} the utilization gap between Attention and FFN pools within a phase, defined as $|U_A - U_F|$, exceeds a threshold, indicating that the current operator-level resource allocation is inefficient.

    \item \underline{\textit{P/D phase skew:}} the utilization gap between prefill and decode phases, defined as $|U_P - U_D|$, exceeds a threshold, indicating that the current phase-level resource allocation no longer matches demand.

    \item \underline{\textit{Workload shift:}} the Kullback-Leibler (KL) divergence between the current input/output length distribution and the workload profile used in the most recent ILP optimization exceeds a threshold, indicating that the current provisioning is based on stale workload assumptions.
\end{itemize}

\begin{figure}[!t]
    \centering
  \centerline{\includegraphics[width=0.48\textwidth]{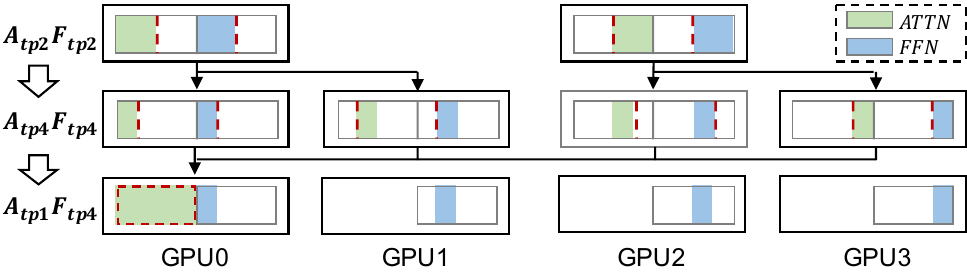}}
    \caption{Independent A/F TP reconfiguration with incremental weight resharding. \AFlex scales $(\mathit{tp}_A,\mathit{tp}_F)$ from $(2,2)$ to $(4,4)$ by reusing resident shards, and then reduces $\mathit{tp}_A$ to $1$.}
    \label{fig:phase-aware-reconfig}
\end{figure}

\noindent\textbf{Resource Orchestrator.}
The Resource Orchestrator jointly provisions the prefill and decode A/F pools. To adapt to workload changes, it periodically selects the number of A/F pairs, TP degrees, and baseline GPU frequencies that minimize energy subject to GPU, memory, latency, and throughput constraints.
Let $c\in\{P,D\}$ denote the prefill or decode phase and $x\in\{A,F\}$ denote the Attention or FFN pool. Phase $c$ contains $k_c$ active A/F pairs, indexed by $i\in\{1,\ldots,k_c\}$. The orchestrator configures pool $x$ in pair $i$ using its TP degree $\mathit{tp}_{cx}^{(i)}$ and baseline frequency $f_{cx}^{(i)}$.
For each candidate configuration and representative workload $r\in\mathcal R_c$, the orchestrator estimates the interval energy $E_x(\mathit{tp},f)$, peak per-GPU memory $\mathrm{Mem}_{cx}^{(i)}(r,\mathit{tp})$, per-layer compute latency $t_{cx}^{(i)}(r)$, and A/F communication latency $t_{\mathrm{comm}}^{c,(i)}(r)$. It also estimates the sustainable throughput $\mu_c(\cdot)$, accounting for pipeline imbalance and communication overhead. We denote the available GPU count by $G$, usable memory per GPU by $C_{\mathrm{GPU}}$, model depth by $L$, workload intensity of phase $c$ by $\lambda_c$, and capacity margin by $\epsilon$.
The orchestrator minimizes total A/F energy while limiting the total allocation to $G$ GPUs and the per-GPU memory footprint to $C_{\mathrm{GPU}}$. It also requires each A/F pair to satisfy its per-layer TTFT or TPOT budget and each phase to sustain $(1+\epsilon)\lambda_c$ throughput. We formulate the optimization in Eq.~\ref{eq:tier1_ilp} as follows:

\begin{figure}[!t]
    \centering
  \centerline{\includegraphics[width=0.48\textwidth]{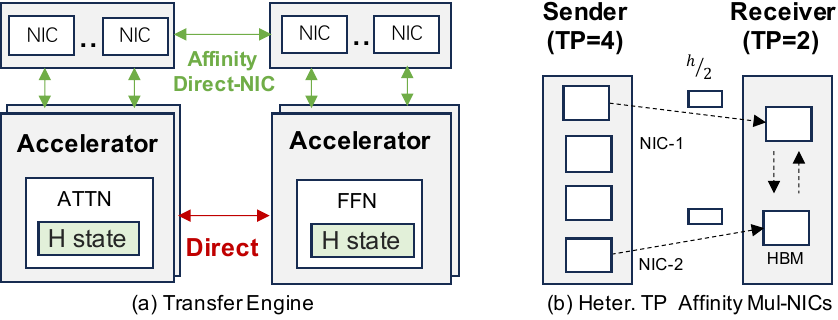}}
    \caption{Placement-aware hidden-state transfer in \AFlexNaN: (a) bidirectional hidden-state transfer across the A/F boundary and (b) NIC-affine transfer across heterogeneous TP groups.}
    \label{fig:transferengine}  
    \vspace{-8pt}
\end{figure}


\begin{subequations}
\label{eq:tier1_ilp}
\begin{align}
\min \quad
&
\sum_{c \in \{P,D\}}
\sum_{i=1}^{k_c}
\left(
E_A(\mathit{tp}_{cA}^{(i)}, f_{cA}^{(i)})
+
E_F(\mathit{tp}_{cF}^{(i)}, f_{cF}^{(i)})
\right)
\label{eq:tier1_obj}
\\[4pt]
\mathrm{s.t.} \quad
&
\sum\nolimits_{c \in \{P,D\}}
\sum\nolimits_{i=1}^{k_c}
\left(
\mathit{tp}_{cA}^{(i)}
+
\mathit{tp}_{cF}^{(i)}
\right)
\le
G
\label{eq:tier1_gpu}
\\
&
\mathrm{Mem}_{cx}^{(i)}(r,\mathit{tp}_{cx}^{(i)})
\le
C_{\mathrm{GPU}}
\label{eq:tier1_mem}
\\
&
t_{PA}^{(i)}(r)
+
t_{PF}^{(i)}(r)
+
t_{\mathrm{comm}}^{P,(i)}(r)
\le
\frac{\mathrm{TTFT}_{\mathrm{SLO}}(r)}{L}
\label{eq:tier1_prefill_slo}
\\
&
t_{DA}^{(i)}(r)
+
t_{DF}^{(i)}(r)
+
t_{\mathrm{comm}}^{D,(i)}(r)
\le
\frac{\mathrm{TPOT}_{\mathrm{SLO}}(r)}{L}
\label{eq:tier1_decode_slo}
\\
&
\sum\nolimits_{i=1}^{k_c}
\mu_c(
\mathit{tp}_{cA}^{(i)},
\mathit{tp}_{cF}^{(i)},
f_{cA}^{(i)},
f_{cF}^{(i)}
)
\ge
(1+\epsilon)\lambda_c
\label{eq:tier1_throughput}
\end{align}
\end{subequations}

To solve this constrained optimization problem efficiently, \AFlex formulates it as an ILP problem over a bounded set of candidate A/F-pair slots. Each active slot uses binary variables to select its TP degree and frequency configuration. Since the profiled latency, memory footprint, energy consumption, and throughput are constants, both the objective and constraints remain linear. \AFlex further reduces the search space by pruning configurations that violate memory or conservative SLO constraints, poorly balanced A/F execution, and Pareto-dominated configurations, enabling the ILP to be solved within each replanning interval.

\noindent\textbf{Runtime Reconfiguration Engine.}
The Runtime Reconfiguration Engine applies new resource allocations without interrupting online serving. It prepares the weight shards and communication groups for the target A/F pools in the background while the current pools continue serving requests, and redirects traffic only after the target pools are ready. To reduce weight-loading overhead, the engine incrementally constructs each target TP layout by first retaining or repartitioning weights on the target GPUs, then fetching missing shards from peer GPUs on the same node, and finally loading the remaining shards from host memory. AFD allows Attention and FFN to be reconfigured independently, but their transition policies differ because only Attention maintains per-sequence KV caches. Active sequences therefore remain on the original Attention pool, while new sequences are routed to the target pool; the original pool is retired after its sequences complete. In contrast, the stateless FFN pool can switch to its target TP layout at the next iteration boundary.

Fig.~\ref{fig:phase-aware-reconfig} illustrates two consecutive reconfigurations, where each A/F layout is represented as (\emph{Attention TP degree}, \emph{FFN TP degree}). \AFlex first scales both pools from $A_{tp2}F_{tp2}$ to $A_{tp4}F_{tp4}$ by repartitioning the existing weight shards, retaining those on existing GPUs, and transferring only those required by newly added GPUs. It then scales down only Attention from TP degree 4 to 1, consolidating its weights onto one GPU while keeping the FFN TP degree at 4.

\subsection{AFD Data Plane}
\label{sec:dataplane}
To minimize the communication overhead introduced by AFD, \AFlex uses an AFD Data Plane that manages request metadata, per-layer hidden-state transfers, and microbatch scheduling. We next describe its placement-aware data path, interleaved A/F pipeline, and DVFS execution model.

\noindent\textbf{Placement-aware A/F Data Path.} As shown in Fig.~\ref{fig:transferengine}(a), the AFD Data Plane provides a unified interface for bidirectional hidden-state transfers at each A/F boundary. The Attention group sends outputs to the FFN group, which returns hidden states for the next layer. \AFlex selects intra-node or cross-node transfer based on group placement.
For cross-node placement, Fig.~\ref{fig:transferengine}(b) shows how \AFlex supports heterogeneous TP configurations without requiring every rank to transfer the full activation. It partitions local ranks into network interface card (NIC) affine groups and selects one representative from each group to send a distinct activation shard. The receiver reassembles and distributes the complete activation through an intra-node collective. 
This design transfers shards in parallel across NICs while keeping inter-node traffic independent of the separately configured ${tp}_A$ and ${tp}_F$.

\noindent\textbf{Interleaved A/F Pipeline.}
AFD forms a two-stage pipeline within each transformer layer, where Attention sends hidden states to FFN and FFN returns its output to the next Attention layer; although standard layer-wise microbatching overlaps the two GPU pools, it creates pipeline bubbles between layers. \AFlex reduces these bubbles through an interleaved schedule that allows each microbatch to proceed once its dependencies are satisfied. To adapt this schedule to dynamic serving loads, \AFlex uses the following mechanisms:

\begin{itemize}[leftmargin=*,itemsep=2pt,parsep=0pt,topsep=2pt]
    \item \underline{\textit{Dynamic microbatch depth $M$:}} \AFlex selects the profiled microbatch depth with the lowest predicted latency for the current workload and A/F configuration.
    \item \underline{\textit{Adaptive request batching:}} Given $M$, \AFlex selects the request batch size with the shortest predicted latency using profiled Attention/FFN latency curves.
\end{itemize}

To account for communication overhead, \AFlex incorporates hidden-state transfer into the A/F execution timeline:
\begin{equation}
\label{eq:afd_comm_model}
t_{\mathrm{layer}} =
\begin{cases}
t_A + t_F + t_{\mathrm{comm}}, & M = 1 \\
\max(t_A,\; t_F) + t_{\mathrm{comm}} / M, & M \ge 2.
\end{cases}
\end{equation}
As shown in Eq.~\eqref{eq:afd_comm_model}, when $M=1$, Attention, transfer, and FFN execute serially. When $M\ge2$, A/F overlap across consecutive microbatches, making the slower stage the compute bottleneck and amortizing transfer cost.

\begin{figure}[!t]
    \centering
    \centerline{\includegraphics[width=0.48\textwidth]{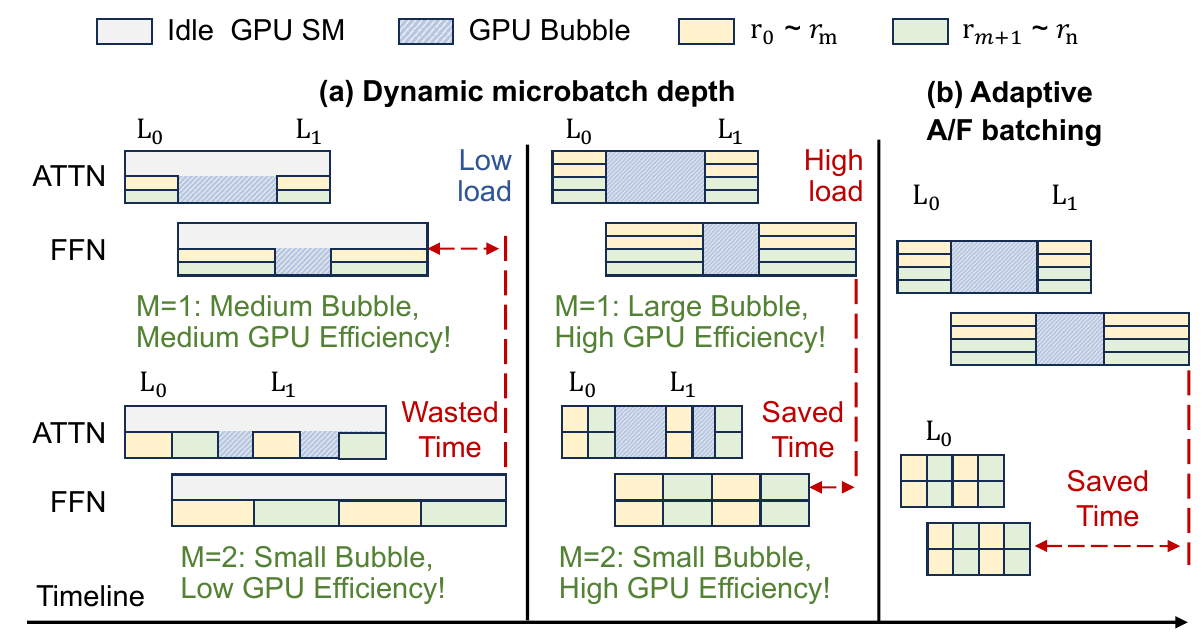}}
    \caption{The interleaved A/F pipeline in \AFlexNaN. \AFlex adapts the microbatch depth (M) to the serving load and balances the work of concurrent A/F stages to reduce GPU idle time.}
    \label{fig:interleaved-af-pipeline}
\end{figure}

Fig.~\ref{fig:interleaved-af-pipeline} illustrates \AFlexNaN's interleaved A/F pipeline. Under low load in Fig.~\ref{fig:interleaved-af-pipeline}(a), the batch is too small to fully utilize the GPU streaming multiprocessors. Although increasing $M$ from 1 to 2 enables A/F overlap, splitting the small batch further reduces GPU utilization, and the resulting computation overhead outweighs the reduction in pipeline bubbles; thus, $M=1$ achieves lower latency. 
Under high load in Fig.~\ref{fig:interleaved-af-pipeline}(a), microbatches are large enough to sustain high GPU utilization. Once $F_{\ell,i}$ completes, the scheduler immediately launches $A_{\ell+1,i}$ without waiting for the remaining microbatches in layer $\ell$, allowing a larger $M$ to reduce pipeline bubbles with minimal computation overhead.
Fig.~\ref{fig:interleaved-af-pipeline}(b) further shows how adaptive request batching balances concurrent $F_{\ell,i}$ and $A_{\ell,i+1}$ stages, reducing GPU idle time.

\subsection{Local DVFS Controller}
The Local DVFS Controller refines the Global Scheduler's plan at millisecond granularity by adjusting GPU frequencies according to the runtime batch size, active request count, and microbatch depth. \AFlex exposes four operator-phase frequency domains, PA, PF, DA, and DF, enabling the controller to exploit the distinct frequency sensitivities of A/F while satisfying TTFT and TPOT SLOs. Because GPU frequency switching incurs non-negligible overhead, the controller accounts for this cost when determining the adjustment interval. We next describe the DVFS Controller from the prefill and decode phases.

\noindent\textbf{Prefill DVFS.}
For prefill, \AFlex minimizes predicted batch energy under the TTFT SLO by selecting a PA/PF frequency pair before launch and keeping it fixed throughout the forward pass.
For the current batch $B$, we define the tightest remaining TTFT budget: $s = \min_{r \in B} \left( \mathrm{TTFT}_{\mathrm{SLO}}(r) - \mathrm{Queue}(r) \right)$, where $\mathrm{Queue}(r)$ is the time from the arrival of request $r$ until batch $B$ begins execution.


\AFlex selects the discrete frequency pairs supported by the hardware based on the TTFT budget $s$, the current PA/PF setting, the Global Scheduler's baseline pair, and the microbatch depth $M$.
The baseline serves as a reference rather than a hard bound: lower frequencies save energy, while higher frequencies are considered when needed to satisfy the SLO. For each candidate, the predictor estimates Attention and FFN latency and energy for the current batch shape (\S\ref{sec:profiling}). The A/F pipeline model then computes the execution time (\S\ref{sec:dataplane}), using the serial bound when $M=1$ and accounting for A/F overlap when $M\ge2$. A candidate is feasible if the predicted time to complete the remaining layers, including frequency-switching overhead, fits within $s$. \AFlex selects the feasible pair with the lowest predicted energy; this search is inexpensive because the hardware exposes only a small set of frequencies.
The selected pair also indicates whether the global plan remains adequate. If the lowest-energy feasible pair exceeds the Global Scheduler's baseline, \AFlex applies it to protect TTFT and reports the escalation to the Workload Monitor (\S\ref{sec:tier1:monitor}). If no pair is feasible, \AFlex falls back to $(f_{\max},f_{\max})$ and reports the infeasibility. Persistent above-baseline or maximum-frequency operation indicates that the global plan may have become stale.

\begin{algorithm}[!t]
\caption{Window-Based Decode DVFS}
\label{alg:decode_dvfs}
\SetKwInOut{Input}{Input}
\SetKwInOut{Output}{Output}

\Input{Decode state $D$; iteration time $t_{\mathrm{iter}}$;
    TPOT SLO $\tau$; batch-size threshold $\delta_b$; \# of microbatches $M$; window size $W$;\\
    controller state $\mathcal{S}=(\mathbf{f},w,b_{\mathrm{last}})$}
\Output{Updated controller state $\mathcal{S}$}

$w\gets w+1$\;

$\Delta_b\gets
|b(D)-b_{\mathrm{last}}|/
\max(b_{\mathrm{last}},1)$\;

\If{
    $t_{\mathrm{iter}}<\tau$
    $\land$
    $\Delta_b<\delta_b$
    $\land$
    $w<W$
}{
    \Return $(\mathbf{f},w,b_{\mathrm{last}})$\;
}

\tcp{Find energy-minimal frequency pair}

$\mathcal{F}_{\tau}\gets
\left\{
\mathbf{f}'\in
\mathcal{F}_{\mathrm{DA}}\times\mathcal{F}_{\mathrm{DF}}
\;\middle|\;
\hat{t}(D,M,\mathbf{f}')\le\tau
\right\}$\;

\If{$\mathcal{F}_{\tau}=\emptyset$}{
    $\mathbf{f}\gets(f_{\max},f_{\max})$\;
    \Return $(\mathbf{f},0,b(D))$\;
}

\tcp{Compute the optimal frequency pair}
$\mathbf{f}^{*}\gets
\mathop{\mathrm{arg\,min}}\nolimits_{\mathbf{f}'\in\mathcal{F}_{\tau}}
\hat{E}(D,\mathbf{f}')$\;

\eIf{
    $\hat{t}(D,M,\mathbf{f})>\tau$
}{
    $\mathbf{f}\gets\mathbf{f}^{*}$\;
}{
    \tcp{Lazily update the frequency}

    \If{
        $\hat{t}(D,M,\mathbf{f}^{*})
        +T_{\mathrm{switch}}/W\le\tau$
    }{
        $\mathbf{f}\gets\mathbf{f}^{*}$\;
    }
}

\Return $(\mathbf{f},0,b(D))$\;
\end{algorithm}

\noindent\textbf{Decode DVFS.}
\label{sec:tier2:decode}
For decode, \AFlex selects DA/DF frequencies to minimize per-token energy under the TPOT SLO. Continuous batching may change the active batch at every iteration boundary, requiring frequent adaptation. However, decode iterations often last only a few milliseconds, making per-iteration frequency switching prohibitively expensive.

The following equation sets the window size using the recent average iteration latency, so each window spans at least $W_{\min}$ iterations and approximately ten switching times:
$W = \max\!\left(W_{\min}, \operatorname{round}\!\left(\kappa T_{\mathrm{switch}} / \bar{t}_{\mathrm{iter}}\right)\right)
$, where $T_{\mathrm{switch}}$ is the frequency-switching latency, $\bar{t}_{\mathrm{iter}}$ is the most recent decode iteration latency, and $\kappa$ controls the amortization horizon.
Algorithm~\ref{alg:decode_dvfs} presents our window-based Decode DVFS algorithm, which selects an energy-efficient DA/DF frequency pair under the TPOT SLO. At each decode iteration, \AFlex updates the window state and computes the relative change in batch size (Lines~1-2). It retains the current pair while the window remains active, the relative batch-size change is below the threshold $\delta_b$, and the iteration time remains within $\tau$; otherwise, it triggers a new search to respond to workload changes or an approaching TPOT limit (Lines~3-4). Once triggered, \AFlex uses the A/F pipeline model to predict the latency and energy of each supported pair, forms the TPOT-feasible set $\mathcal{F}_{\tau}$, and selects its lowest-energy pair $\mathbf{f}^{*}$ (Lines~5 and~9). If $\mathcal{F}_{\tau}$ is empty, it falls back to the maximum-frequency pair for best-effort SLO protection (Lines~6-8). \AFlex applies $\mathbf{f}^{*}$ immediately if the current pair can no longer satisfy TPOT (Lines~10-11); otherwise, it switches lazily only when the predicted latency, including the amortized switching delay $T_{\mathrm{switch}}/W$, remains within $\tau$ (Lines~13-14). Finally, \AFlex resets the window state for the next decision (Line~15).

\section{Implementation}
\label{sec:implementation}

We implement \AFlex atop SGLang~\cite{sglang}. \AFlex consists of a profiling module, a scheduling control plane, and an AFD data plane, comprising approximately 15K SLOC of Python and 1K SLOC of C++.

\noindent\textbf{Profiling module.}
We collect A/F latency and energy profiles by sweeping TP degree, GPU frequency, batch size, and sequence length, measuring GPU energy with NVML~\cite{NVML}. 
Based on these profiles, we use scikit-learn~\cite{scikit-learn} to train separate latency and energy predictors for each A/F role, as well as an interleaved pipeline model for decode. The trained models are serialized and loaded at startup, enabling low-overhead queries by the two-level control plane.

\noindent\textbf{Scheduling control plane.} 
The scheduling control plane comprises the Global Resource Controller and the Local DVFS Controller.
The Global Resource Controller is implemented as a lightweight background process comprising a Workload Monitor, a Resource Orchestrator, and a Reconfiguration Engine.
The Workload Monitor tracks SLO violations, resource utilization, tail latency, and request load. It triggers replanning when it detects sustained SLO violations, A/F imbalance, P/D phase skew, or workload shifts.
Using the profiling models, the Resource Orchestrator selects the minimum-energy A/F configuration satisfying memory and SLO constraints. The Reconfiguration Engine updates A/F slots, reinitializes the NCCL communicators for affected TP groups, and loads the required weight shards, while frequency-only changes are applied in place. 
At runtime, the Local DVFS Controller independently adjusts four operator pool frequencies via NVML~\cite{NVML}.

\noindent\textbf{AFD data plane.}
Each worker initially loads only its role-specific and shared weights and skips the inactive operator during execution. The Reconfiguration Engine loads missing weight shards on demand when the target A/F configuration requires a worker to assume a different role or TP degree.
The PA, PF, DA, and DF pools are implemented as separate SGLang scheduler processes that coordinate via ZMQ~\cite{ZeroMQ}.
Hidden states are transferred across nodes through UCX-Py~\cite{ucx-py} over GPUDirect RDMA and within a node through CUDA IPC and \texttt{cudaMemcpyPeer} over NVLink.
To make inter-node traffic independent of the TP degree, sender ranks partition the hidden states into disjoint token shards and transmit one shard independently. 
The receiving ranks then perform \texttt{all\_gather} within each node to reconstruct the complete hidden states.

\section{Evaluation}
\subsection{Experiment Setup}

\noindent\textbf{Hardware and Software Configuration.}
We conduct the main experiments on two NVIDIA A800 servers and scale to four servers for the scalability study. 
Each server has eight NVIDIA A800-80GB GPUs connected by 400\,GB/s NVLink, an 800\,Gbps inter-node network, a 128-core 2.50\,GHz Intel Xeon Platinum CPU, and 1\,TB DRAM.
The software stack uses Ubuntu 22.04 LTS and CUDA 12.4.

\noindent\textbf{Models and Workloads.}
We evaluate \AFlex using Qwen3-32B~\cite{qwen3technicalreport} and Mixtral-8$\times$7B~\cite{Mixtral} as representative dense and mixture-of-experts (MoE) models, respectively. These models exhibit different computation and communication characteristics, allowing us to evaluate \AFlex across model architectures.

We use two representative production traces from Azure~\cite{azure-datasets}, namely \textit{Coding} and \textit{Conversation}, for the end-to-end evaluation. The \textit{Coding} trace contains requests with relatively long input contexts and short outputs, whereas the \textit{Conversation} trace has shorter input contexts and more variable output lengths. In addition, we construct four controlled workloads that cover different combinations of prefill and decode demands: \textit{Question Answering} (QA, 128 input and 64 output tokens), \textit{Retrieval-Augmented Generation} (RAG, 4096 input and 64 output tokens), \textit{Chatbot} (128 input and 1024 output tokens), and \textit{Summary} (4096 input and 1024 output tokens). These workloads enable a systematic evaluation of \AFlex under diverse input/output characteristics.

\noindent\textbf{Metrics and Methodology.}
To quantify the energy savings achieved by \AFlex while satisfying latency SLOs, we use energy per token, defined as the ratio of energy consumed by all GPUs to the total number of input and output tokens, as the primary energy-efficiency metric and report P90 TTFT and TPOT to assess SLO compliance. We measure GPU energy using NVML, calculate energy per token, and vary request rates to evaluate the systems under different serving loads. Unless otherwise specified, we set the P90 latency SLOs for TTFT and TPOT to 400\,ms and 120\,ms, respectively. 

\noindent\textbf{Baseline Systems.}
We compare \AFlex against four baselines covering colocated serving, P/D disaggregation, and different GPU frequency-control granularities:
\begin{itemize}[leftmargin=*,itemsep=2pt,parsep=0pt,topsep=2pt]
    \item \textit{SGLang} \cite{sglang}: a P/D-colocated serving system without adaptive GPU frequency control.
    \item \textit{DynamoLLM} \cite{stojkovic2025dynamollm}: a DVFS-enabled P/D-colocated serving system that dynamically adjusts GPU frequencies.
    \item \textit{DistServe} \cite{zhong2024distserve}: a P/D-disaggregated serving system without adaptive GPU frequency control.
    \item \textit{BiScale} \cite{basit2026biscale}: a DVFS-enabled P/D-disaggregated serving system with phase-level GPU frequency control.
\end{itemize}

\subsection{End-to-End Energy Efficiency under Production traces}
\label{subsec:e2e}
\begin{figure}[t]
  \centering
  \includegraphics[width=0.48\textwidth]{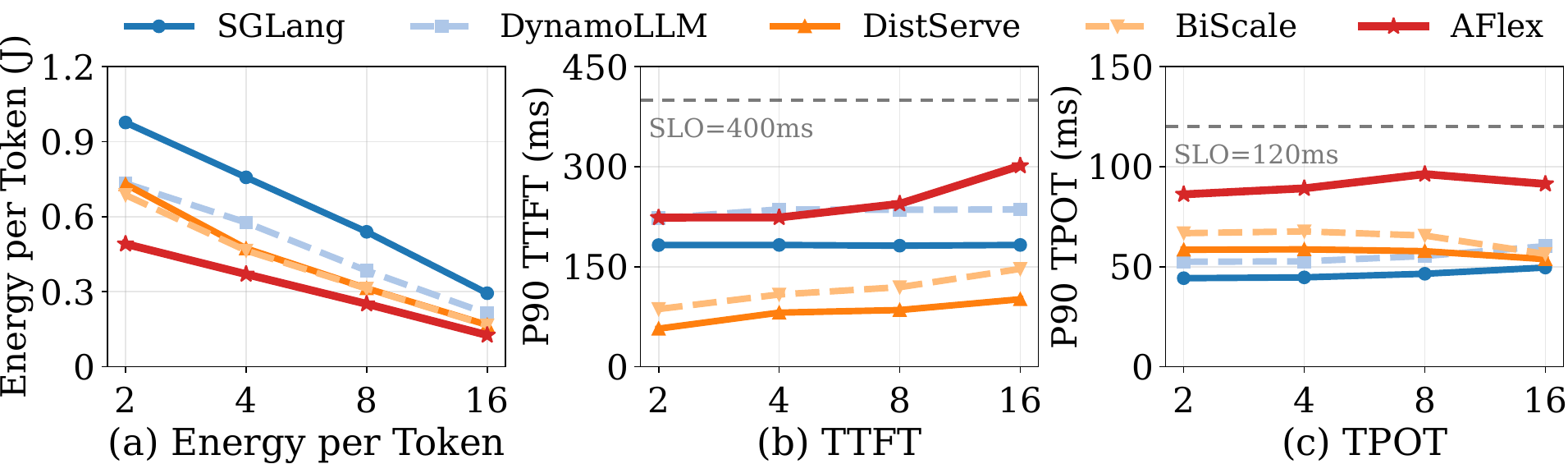}
  \caption{End-to-end performance under the Azure Conversation trace: (a) Energy per token, (b) P90 TTFT, and (c) P90 TPOT as the RPS increases. 
  }
  \label{fig:e2e1}
\end{figure}

\begin{figure}[t]
  \centering
  \includegraphics[width=0.48\textwidth]{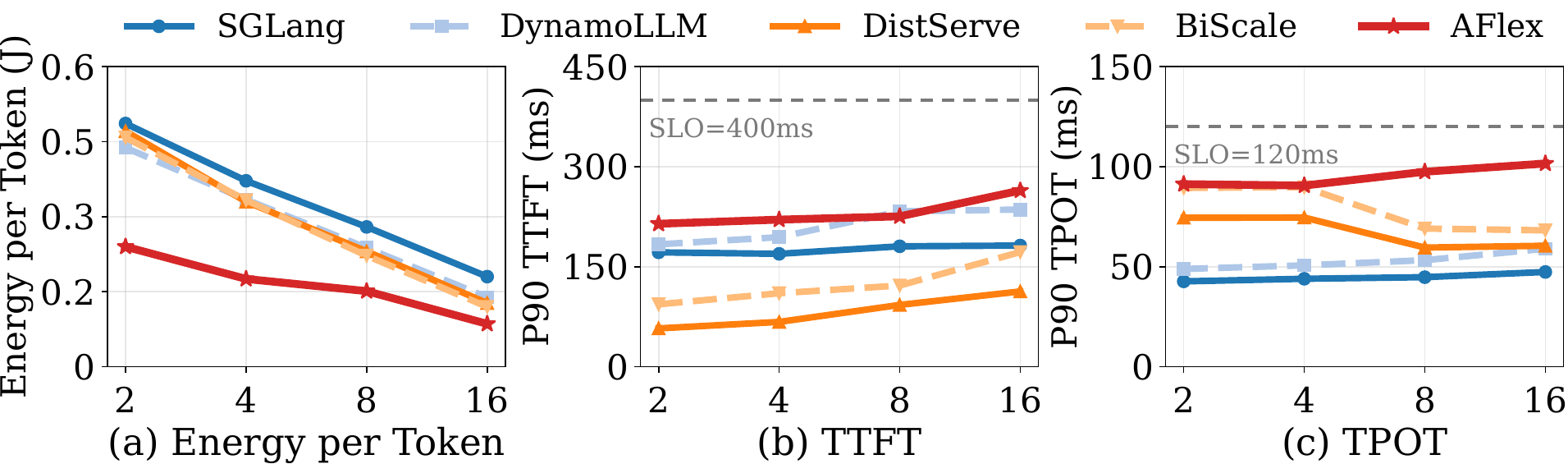}
  \caption{End-to-end performance under the Azure Coding trace: (a) Energy per token, (b) P90 TTFT, and (c) P90 TPOT as the RPS increases.}
  \label{fig:e2e2}
\end{figure}

We first evaluate \AFlex using production traces at 2, 4, 8, and 16 requests per second (RPS).
Figs.~\ref{fig:e2e1} and~\ref{fig:e2e2} report energy per token, P90 TTFT, and P90 TPOT.

\noindent\textbf{Coarse-grained DVFS provides limited energy savings.}
Fig.~\ref{fig:e2e1}(a) shows the energy per token under the \textit{Conversation} trace. Compared with SGLang, DynamoLLM reduces energy per token by at most 28.8\%, and the reduction decreases to 27.6\% at 16 requests/s because its shared frequency setting cannot capture the heterogeneous frequency sensitivities of different execution phases. BiScale exhibits a similar limitation, reducing energy by only 0.3\%--6.0\% relative to DistServe because all stages within each phase still share the same frequency. In contrast, DistServe reduces energy per token by up to 43.0\% over SGLang despite not using adaptive DVFS, suggesting that P/D disaggregation exposes additional opportunities for energy-efficient resource allocation.

\noindent\textbf{Fine-grained disaggregation and DVFS are complementary.}
By combining AFD with fine-grained resource provisioning and DVFS, \AFlex independently adapts the GPU allocation and frequency of PA, PF, DA, and DF. Consequently, it achieves the lowest energy per token at every RPS under the \textit{Conversation} trace, outperforming the best-performing baseline by up to 28.5\%. Fig.~\ref{fig:e2e2}(a) shows a similar trend under the \textit{Coding} trace. Although its longer input contexts introduce substantially more prefill computation and increase energy consumption across all systems, \AFlex consistently achieves the lowest energy per token. Across the two production traces, \AFlex reduces energy per token by up to 49\% over DistServe and 48\% over DynamoLLM.

\noindent\textbf{\AFlex preserves latency SLOs while reducing energy.}
Fig.~\ref{fig:e2e1}(b) shows the corresponding P90 TTFT. P/D disaggregation allows DistServe to substantially reduce TTFT compared with SGLang. In contrast, DynamoLLM and BiScale increase TTFT relative to their respective non-DVFS baselines due to lower GPU frequencies. Despite achieving the largest energy savings, \AFlex keeps P90 TTFT below 302\,ms at every RPS. Fig.~\ref{fig:e2e1}(c) further shows that \AFlex maintains P90 TPOT below 97\,ms, satisfying the TPOT SLO across all evaluated loads. Under the \textit{Coding} trace in Fig.~\ref{fig:e2e2}, \AFlex similarly keeps P90 TTFT and TPOT below 265\,ms and 102\,ms, respectively. Overall, while satisfying both latency SLOs across all evaluated RPS, \AFlex reduces energy per token by 19.6\%--46.8\% relative to the best-performing baseline.

\begin{figure}[t]
    \centering
    \includegraphics[width=\linewidth]{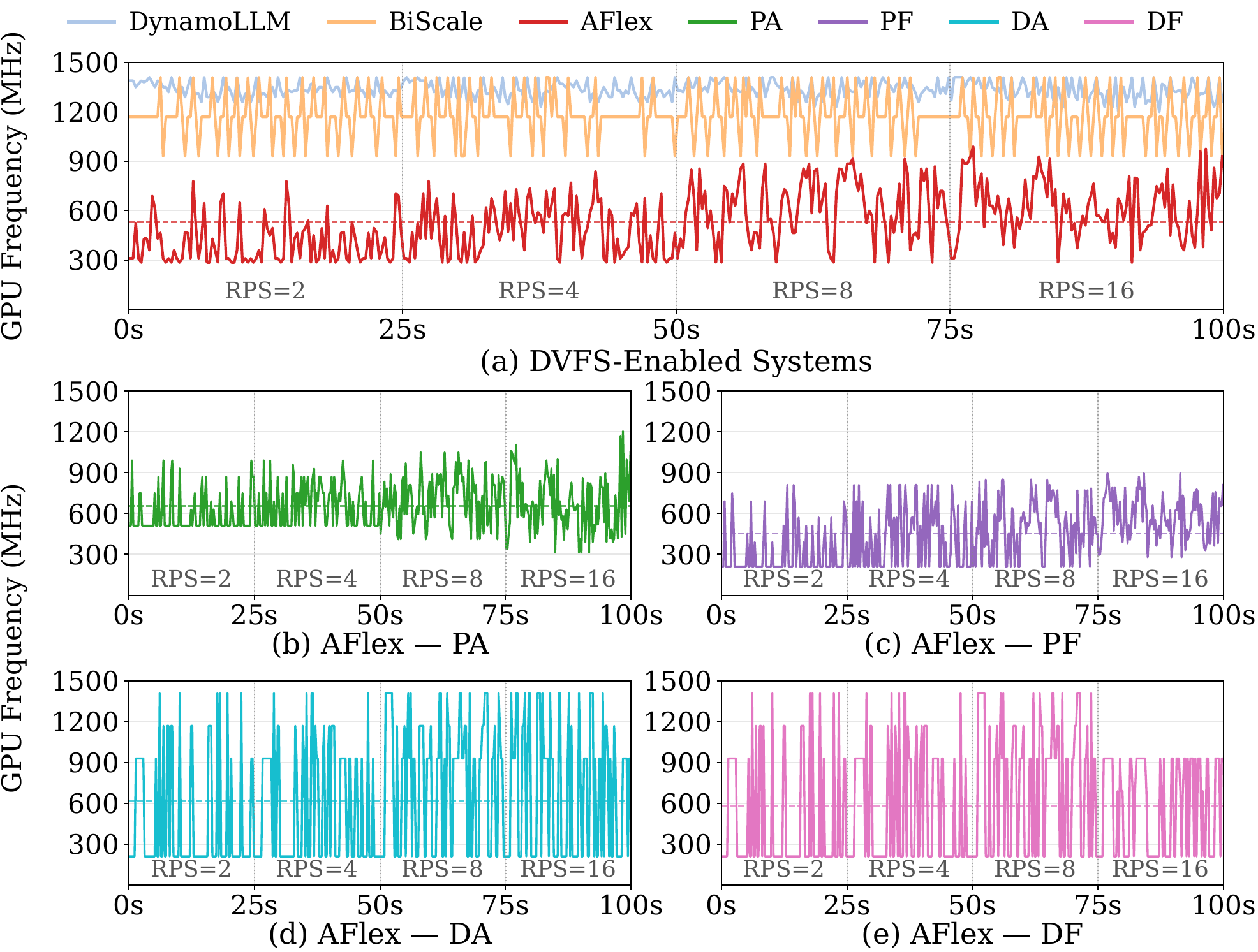}
    \caption{GPU frequency timelines for Coding trace on 16 GPUs across two nodes.}
    \label{fig:e2e3}
\end{figure}

\noindent\textbf{Fine-grained control unlocks greater frequency-reduction opportunities.}
We next examine GPU-frequency timelines under the \textit{Coding} trace as RPS increases. Fig.~\ref{fig:e2e3}~(a) compares \AFlexNaN's frequency timeline with those of the two DVFS-enabled baselines. SGLang and DistServe are omitted because they consistently run all GPUs at the maximum frequency of 1410\,MHz. The P/D-colocated DynamoLLM forces both phases to share a frequency high enough for compute-intensive prefill, leaving little room for frequency reduction. BiScale disaggregates P/D and averages 1357\,MHz and 966\,MHz for prefill and decode, respectively, but its phase-level control cannot exploit operator-specific slack for further frequency reduction. In contrast, \AFlex reduces the overall average GPU frequency to 530\,MHz, as indicated by the red dashed line.


Figs.~\ref{fig:e2e3}(b)--(e) reveal the reason: PA generally operates at a higher frequency because long inputs
disproportionately increase its latency relative to PF, making it the bottleneck of the prefill pipeline. \AFlex exploits the resulting pipeline slack, enabling PF to run at a lower frequency without reducing pipeline throughput.
Meanwhile, the short outputs tend to keep decode batches small under low load, reducing the frequency sensitivity of DA and DF. By tuning each stage independently, \AFlex reduces the average frequencies of PA, PF, DA, and DF to 656, 452, 616, and 580\,MHz, respectively.

\begin{figure}[t]
    \centering
    \includegraphics[width=\linewidth]{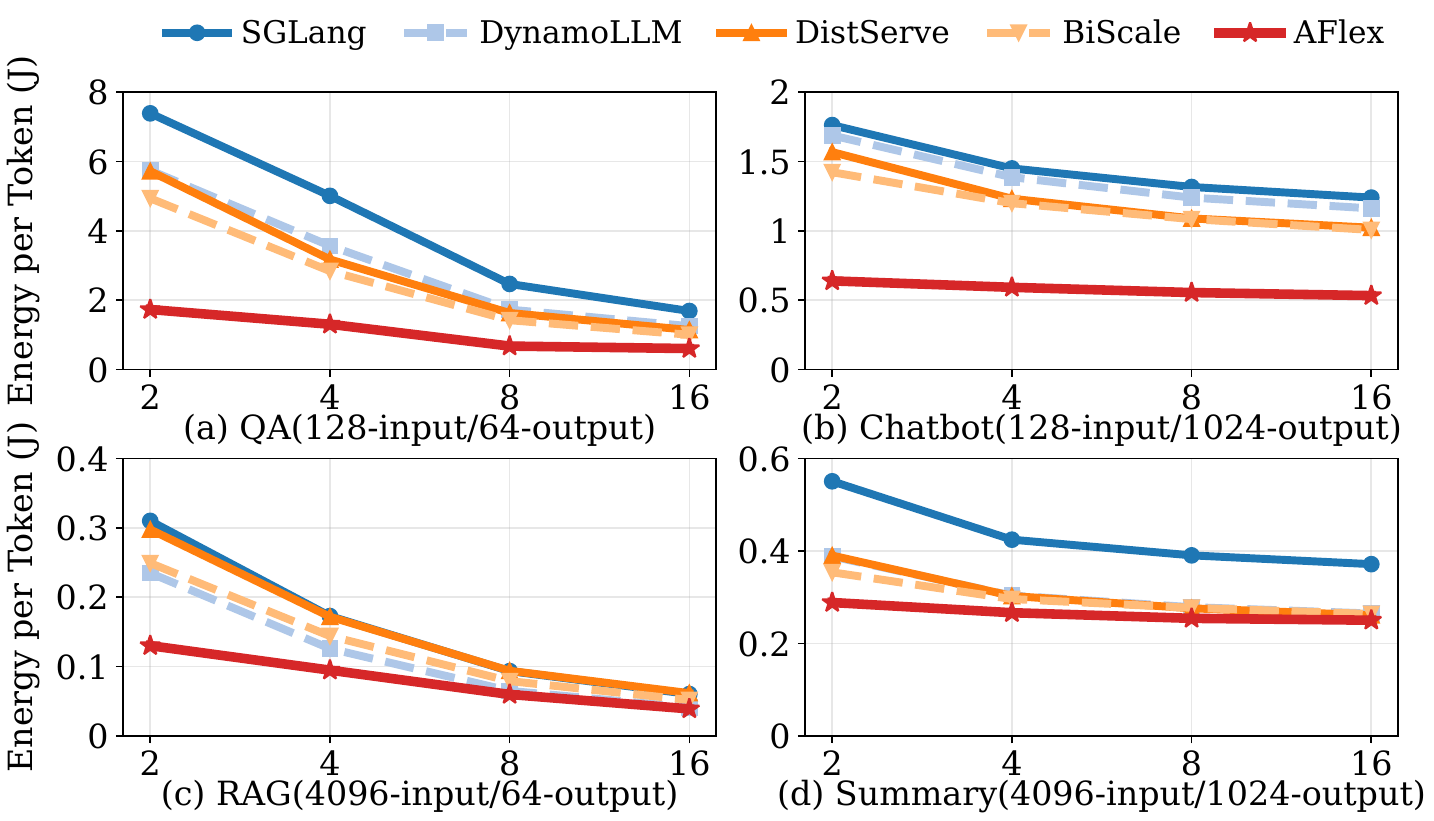}
    \caption{Energy efficiency: controlled workloads, 2--16 RPS.
    }
    \label{fig:micro_all}    
\end{figure}

\subsection{Impact of Workload Characteristics}
To evaluate how \AFlex balances energy consumption and latency under different workload characteristics, we construct four controlled workloads with varying input and output lengths. As shown in Fig.~\ref{fig:micro_all}, we start with the lightweight \textit{QA} (prefill-light, decode-light) workload, which contains 128 input and 64 output tokens. \textit{Chatbot} (prefill-light, decode-heavy) increases only the output length to 1024 tokens to create a decode-heavy workload, while \textit{RAG} (prefill-heavy, decode-light)  increases only the input length to 4096 tokens to create a prefill-heavy workload.
\textit{Summary} (prefill-heavy, decode-heavy) combines both changes, stressing the prefill and decode phases simultaneously. All evaluated settings satisfy both the TTFT and TPOT SLOs.

\noindent\textbf{\AFlex preserves energy efficiency under light load.}
We observe that all baselines consume substantially more energy per token at low RPS under the QA workload. As shown in Fig.~\ref{fig:micro_all}(a), the baselines consume 4.9--7.4\,J per token at 2 RPS, despite the low prefill and decode demands. These values are 4.4--5.0$\times$ those at 16 RPS, indicating that their coarse-grained controls cannot efficiently adapt to light-load execution. In contrast, \AFlex consumes only 1.7\,J per token at 2 RPS. Across the four workloads, its energy per token at 2 RPS is only 1.2--3.3$\times$ that at 16 RPS, indicating that \AFlex better preserves energy efficiency as the RPS decreases.

\noindent\textbf{Asymmetric workloads require fine-grained joint control.}
Relative to QA, Chatbot and RAG increase the pressure on the decode and prefill phases, respectively. Under the decode-heavy Chatbot workload in Fig.~\ref{fig:micro_all}(b), coarse-grained DVFS reduces energy by only 4--9\% when comparing DynamoLLM with SGLang and BiScale with DistServe. Fig.~\ref{fig:micro_all}(c) similarly shows that P/D disaggregation alone provides limited benefits under the prefill-heavy RAG workload. When one phase dominates execution, coarse-grained controls must select configurations that accommodate its bottleneck operator and cannot exploit the remaining optimization opportunities in other operators. By independently provisioning and scaling PA, PF, DA, and DF, \AFlex consistently achieves the lowest energy per token, reducing it by up to 55\% and 45\% over the best-performing baseline under Chatbot and RAG, respectively.

\noindent\textbf{\AFlex remains efficient under joint phase pressure.}
The Summary workload in Fig.~\ref{fig:micro_all}(d) combines a long input with a long output, increasing both prefill and decode demands. Although the energy per token of all systems decreases as the RPS increases, \AFlex remains the most energy-efficient across all evaluated rates. Compared with the best-performing baseline, \AFlex reduces energy per token by 4--19\% while satisfying the TTFT and TPOT SLOs.

\subsection{Ablation Study}

We further evaluate \AFlex along three dimensions: energy efficiency across cluster sizes and model architectures, the contributions of its key mechanisms, and system overheads.

\begin{figure}[t]
    \centering
    \includegraphics[width=\linewidth]{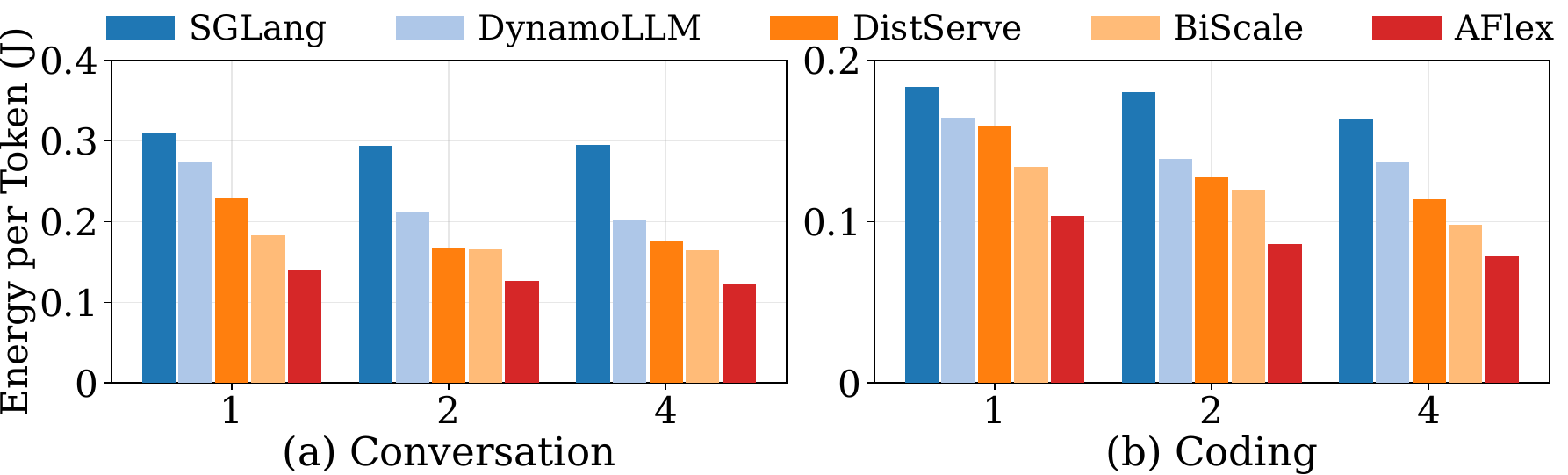}
    \caption{Energy per token scaling from one to four nodes at 8 RPS per node (a) Conversation  (b) Coding.}
    \label{fig:node-scalability}    
\end{figure}

\noindent\textbf{Scale-out Scalability.}
We scale the cluster from 1 to 4 nodes at a fixed 8~RPS per node (8--32~RPS aggregate).
Fig.~\ref{fig:node-scalability} shows that although the baselines modestly reduce energy per token as the cluster scales, a substantial gap from \AFlex persists, indicating considerable room for operator-level energy optimization. Across both workloads and all cluster sizes, \AFlex reduces energy per token by at least 20.2\% relative to the best-performing baseline, demonstrating that its energy-saving advantage persists with scale.


\begin{figure}[t]
    \centering
    \includegraphics[width=\linewidth]{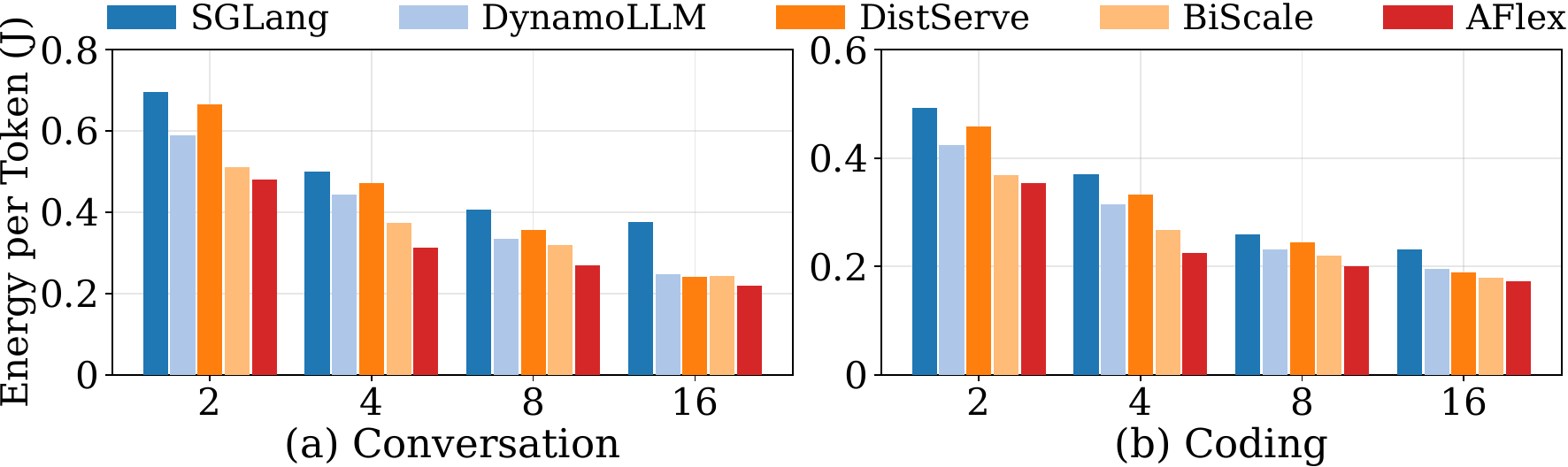}
    \caption{Energy per token of MoE Mixtral-8$\times$7B under 2--16 RPS on two nodes: (a) Conversation and (b) Coding.}
    \label{fig:model-scalability}
\end{figure}

\noindent\textbf{MoE Model Generality.}
We evaluate \AFlex on MoE Mixtral-8$\times$7B using the same experimental setup as \cref{subsec:e2e}. Compared with the dense-model results in Figs.~\ref{fig:e2e1} and~\ref{fig:e2e2}, DistServe's energy savings over SGLang diminish because sparse expert execution changes FFN computation and communication without similarly affecting Attention, creating intra-phase A/F imbalance that P/D disaggregation cannot address. As Fig.~\ref{fig:model-scalability} shows, \AFlex addresses this imbalance through independent A/F provisioning and frequency control, achieving the lowest energy per token under both workloads and reducing it by up to 41.4\% over the best-performing baseline.


\begin{figure}[t]
\centering
\includegraphics[width=\linewidth]{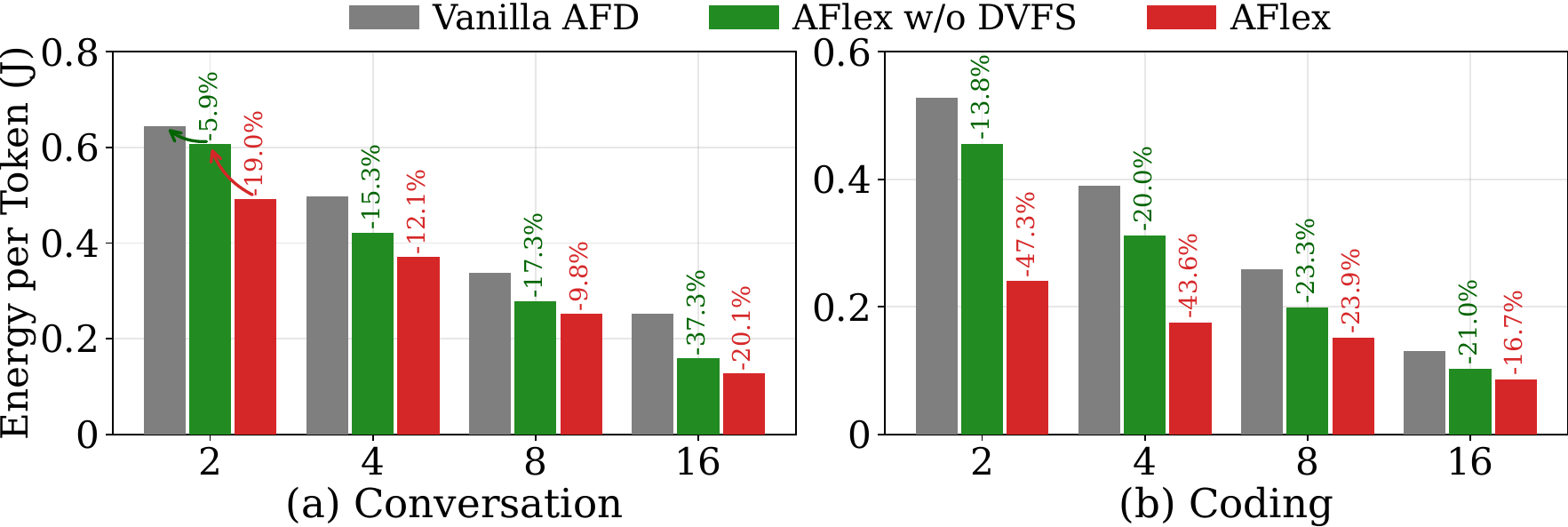}
\caption{
Benefits of Global Scheduler and Local DVFS.
}\label{fig:performance-breakdown}
\end{figure}

\noindent\textbf{Benefit Breakdown.}
Starting from a Vanilla AFD system, we incrementally enable the Global Scheduler (\AFlex w/o DVFS) and Local DVFS Controller (\AFlexNaN) on two 8-GPU nodes. As Fig.~\ref{fig:performance-breakdown} shows, under \textit{Conversation}, variable output lengths create cross-stage resource imbalance that becomes more pronounced as the RPS increases. Consequently, the scheduler's energy savings grow from 5.9\% to 37.3\%, while local DVFS provides an additional reduction of up to 20.1\%. Under \textit{Coding}, long contexts make prefill more compute-intensive, while low RPS leave sufficient latency slack for frequency reduction. Local DVFS exploits this slack to reduce energy per token by up to 47.3\%.

\begin{table}[t]
    \centering
    \caption{Benefits of dynamic microbatch depth $M$ and adaptive request batching during decode. Low and high RPS use batch sizes of 2 and 384; the latter use equal ($192+192$) or adaptive ($128+256$). Layer time denotes per-layer latency.}
    \label{tab:adaptive-batching}
    \begingroup
    \small
    \setlength{\tabcolsep}{2.5pt}
    \renewcommand{\arraystretch}{0.85}
    \setlength{\aboverulesep}{1pt}
    \setlength{\belowrulesep}{1pt}
    \begin{tabular*}{\columnwidth}{@{\extracolsep{\fill}} l l c c c @{}}
        \toprule
        Optimization
        & Workload
        & Configuration
        & Layer time
        & Bubble  \\
        \midrule
        \multirow[c]{2}{*}{Dynamic depth}
        & Low RPS
        & $M: 2 \rightarrow 1$
        & -37.9\%
        & -8.5\% \\
        & High RPS
        & $M: 1 \rightarrow 2$
        & -20.0\%
        & -65.1\% \\
        \midrule
        Request batching
        & High RPS
        & Equal $\rightarrow$ Adaptive
        & -19.9\%
        & -16.1\% \\
        \bottomrule
    \end{tabular*}
    \endgroup
    \vspace{8pt}
\end{table}

\noindent\textbf{Benefits of dynamic microbatch depth and adaptive request batching.}
We evaluate the effectiveness of dynamic microbatch depth and adaptive request batching (see details in \cref{sec:dataplane}) during decode. As shown in Table~\ref{tab:adaptive-batching}, under low RPS load, \AFlex retains $M=1$ to avoid splitting an already small batch, reducing pipeline bubble time by 8.5\% and per-layer execution time by 37.9\% compared with $M=2$. Under high RPS load, \AFlex instead selects $M=2$ to increase A/F overlap, reducing bubble time by 65.1\% and per-layer execution time by 20.0\% compared with $M=1$. In addition, adaptive request batching better balances the A/F stages, further reducing bubble time by 16.1\% and per-layer execution time by 19.9\% compared with equal partitioning.




\noindent\textbf{Planning and Reconfiguration Overheads.}
We then evaluate whether \AFlex is lightweight by measuring its global resource planning and TP reconfiguration overheads.
As shown in Table~\ref{tab:tier1-solver-breakdown}, the ILP solver takes 0.49--0.57\,s as the cluster scales from 8 to 32 GPUs, accounting for less than 0.2\% of the 5\,min scheduling window. Table~\ref{tab:reshard-vs-baseline} shows that incremental resharding reduces reconfiguration time by 52.2\% when increasing the TP degree from 2 to 4 and by 72.3\% when decreasing it from 4 to 1. These reductions come from reusing resident weight shards, transferring only the missing shards over intra-node NVLink, and preparing the target configuration asynchronously while the current instance continues serving requests. Overall, \AFlex incurs negligible planning overhead ($<1\%$) and substantially reduces reconfiguration cost.

\noindent\textbf{Predictor Accuracy.}
Finally, we evaluate \AFlexNaN's predictor using the $R^2$ score and Mean Absolute Percentage Error (MAPE) under 90/10 and 10/90 train/test splits. Across both splits, all prediction tasks achieve $R^2 \geq 0.97$. Prefill MAPE remains below 11\%, while decode MAPE stays around 2\%. These results show that \AFlex can make reliable global scheduling decisions based on accurate performance estimates, even with sparse training sets.

\begin{table}[!t]
\centering
\caption{
  Average ILP latency breakdown (scheduling window = 5\,min).
  All values are in milliseconds.
}
\label{tab:tier1-solver-breakdown}
\begingroup
\small
\setlength{\tabcolsep}{2.5pt}
\renewcommand{\arraystretch}{0.9}
\setlength{\aboverulesep}{1pt}
\setlength{\belowrulesep}{1pt}
\begin{tabular*}{\columnwidth}{@{\extracolsep{\fill}} l c c c c c c @{}}
\toprule
& \multicolumn{3}{c}{Conversation}
& \multicolumn{3}{c}{Coding} \\
\cmidrule(lr){2-4} \cmidrule(lr){5-7}
& $G{=}8$ & $G{=}16$ & $G{=}32$
& $G{=}8$ & $G{=}16$ & $G{=}32$ \\
\midrule
Search
& 490.33 & 532.88 & 479.13
& 570.42 & 544.75 & 510.62 \\
Other
& 1.12 & 3.47 & 7.50
& 1.35 & 3.84 & 7.83 \\
Total
& 491.45 & 536.35 & 486.63
& 571.77 & 548.59 & 518.45 \\
\textbf{Overhead (\%)}
& \textbf{0.16} & \textbf{0.18} & \textbf{0.16}
& \textbf{0.19} & \textbf{0.18} & \textbf{0.17} \\
\bottomrule
\end{tabular*}
\endgroup
\end{table}

\begin{table}[!t]
\centering
\caption{Reconfiguration overhead (s): NCCL setup (Comm.), weight/KV cache init (W/KV), and RDMA init.}
\label{tab:reshard-vs-baseline}
\begingroup
\footnotesize
\setlength{\tabcolsep}{0.6pt}
\renewcommand{\arraystretch}{0.9}
\setlength{\aboverulesep}{1pt}
\setlength{\belowrulesep}{1pt}
\begin{tabular*}{\columnwidth}{@{\extracolsep{\fill}} l c c c c c c c c @{}}
\toprule
& \multicolumn{4}{c}{TP2$\!\to\!$TP4}
& \multicolumn{4}{c}{TP4$\!\to\!$TP1} \\
\cmidrule(lr){2-5}
\cmidrule(lr){6-9}
& Comm. & W/KV & RDMA & Total
& Comm. & W/KV & RDMA & Total \\
\midrule
Base.
& 1.50 & 5.40 & 1.40 & 8.30
& 0.16 & 11.10 & 1.30 & 12.56 \\
\AFlex
& 0.71 & 2.16 & 1.10 & \textbf{3.97}
& 0.71 & 1.48 & 1.29 & \textbf{3.48} \\
\midrule
\textbf{Reduction}
& \multicolumn{3}{c}{--} & \textbf{52.2\%}
& \multicolumn{3}{c}{--} & \textbf{72.3\%} \\
\bottomrule
\end{tabular*}
\endgroup
\vspace{8pt}
\end{table}

\section{Related Work}
\noindent\textbf{LLM serving and disaggregated inference.}
Prior LLM serving systems improve the execution efficiency of each serving instance through continuous batching~\cite{yu2022orca}, memory optimization~\cite{agrawal2023sarathi, flexgen-pmlr23}, KV cache management~\cite{qin2024mooncake, hu2024memserve, liu2025lmcache}, and efficient attention~\cite{dao2022flashattention, dao2024flashattention, flashdecoding++-2023, zadouri2026flashattention}. These techniques optimize the execution path of LLM inference. Disaggregated LLM serving systems distribute inference across GPU pools at phase (P/D) or operator (A/F) granularity to improve latency, throughput, or resource utilization~\cite{zhong2024distserve, patel2023splitwise, tetriserve-2024, zhu2025megascale}. \AFlex is orthogonal and complementary: it targets energy reduction through DVFS while preserving SLOs.

\noindent\textbf{Power Management and Energy-efficient Serving.}
Power management is a common approach to improve datacenter energy efficiency while preserving SLOs~\cite{hsu2018smoothoperator, patel2024characterizing}. Recent studies have characterized the energy, power, and carbon cost of LLM inference~\cite{chien2023reducing, stojkovic2024towards}, prompting efforts to reduce serving energy through placement, parallelism selection, and GPU frequency adjustment. DynamoLLM~\cite{stojkovic2025dynamollm} groups requests by input/output length and optimizes parallelism configurations within each pool, while throttLL'eM~\cite{kakolyris2025throttll} applies predictive GPU throttling based on latency slack. GreenLLM~\cite{greenllm} and BiScale~\cite{basit2026biscale} study the energy-efficient P/D-disaggregated serving. These systems manage energy at the cluster, instance, pool, or phase granularity. In contrast, \AFlex targets AFD serving and jointly controls A/F provisioning and PA/PF/DA/DF frequencies under TTFT/TPOT constraints.

\label{sec:related}

\section{Conclusion}
We present \AFlexNaN, an energy-efficient framework for AFD LLM serving. \AFlex combines an interleaved A/F pipeline with dynamic microbatch depth and request batch size, together with a two-level control plane consisting of a global scheduler and local DVFS controller. 
\AFlex reduces energy per token by up to 49\% over state-of-the-art disaggregated serving and 48\% over frequency-scaling systems while satisfying TTFT and TPOT SLOs.

\bibliographystyle{IEEEtranS}
\bibliography{refs}

\end{document}